\documentclass[fleqn,useAMS,usenatbib]{mnras}
\usepackage{amssymb}     \usepackage{amsmath}    \usepackage{amsfonts}
\usepackage{xspace}    \usepackage{longtable}   \usepackage{rotating} 
\usepackage{lscape}
\usepackage{times}
\usepackage{soul}
\usepackage{xcolor}
\usepackage[countmax]{subfloat}

\title[Radio Galaxy Zoo DR1]
  {Radio Galaxy Zoo Data Release 1: 100,185 radio source classifications from the FIRST and ATLAS surveys}
\author[O.\ I.\ Wong et al.]{O.~Ivy~Wong,$^{1,2}$\thanks{E-mail:ivy.wong@csiro.au} A.~F.~Garon,$^{3}$ M.~J.~Alger,$^{4,5}$ L.~Rudnick,$^{3}$ S.~S.~Shabala,$^{6}$ K.~W.~Willett,$^{3}$   
\newauthor  J.~K.~Banfield,$^{4}$ H.~Andernach,$^{7}$ R.~P.~Norris,$^{8,9}$  J.~Swan,$^{6}$  M.~J.~Hardcastle,$^{10}$  C.~J.~Lintott,$^{11}$ 
\newauthor S.~V.~White,$^{12,13}$   N.~Seymour,$^{14}$ A.~D.~Kapi\'{n}ska,$^{15}$   H.~Tang,$^{16}$ B.~D.~Simmons$^{17}$ \& K. Schawinski$^{18,19}$
\\
$^1$ ATNF, CSIRO Space \& Astronomy, PO Box 1130, Bentley, Western Australia 6102, Australia \\
$^2$ International Centre for Radio Astronomy Research, The University of Western Australia M468, 35 Stirling Highway, Crawley, WA 6009, Australia\\
$^3$ Minnesota Institute for Astrophysics, University of Minnesota, 116 Church St. SE, Minneapolis, MN 55455, USA \\
$^4$ Research School of Astronomy and Astrophysics, Australian National University, Weston Creek, ACT 2611, Australia \\
$^5$ Data61, CSIRO, Canberra, ACT 2601, Australia\\
$^{6}$School of Physical Sciences, University of Tasmania, Private Bag 37, Hobart, Tasmania 7001, Australia\\
$^7$ Th\"uringer Landessternwarte, Sternwarte 5, D-07778 Tautenburg, Germany; on 
  leave of absence from Departamento de Astronom\'{i}a,\\
   DCNE, Universidad de Guanajuato, Callej\'on de Jalisco s/n, CP 36023, Guanajuato, Gto., Mexico\\
$^{8}$ Western Sydney University, Locked Bag 1797, Penrith South, NSW 1797, Australia\\
$^{9}$ ATNF, CSIRO Space \& Astronomy, PO Box 76, Epping, NSW 1710, Australia\\
$^{10}$ Department of Physics, Astronomy \& Mathematics, University of Hertfordshire, College Lane, Hatfield AL10 9AB, UK\\
$^{11}$ Oxford Astrophysics, Denys Wilkinson Building, Keble Road, Oxford OX1 3RH, UK\\
$^{12}$ South African Astronomical Observatory, P.O. Box 9, Observatory 7935, South Africa\\
$^{13}$ Rhodes University, Rhodes Centre for Radio Astronomy Techniques \& Technologies, P.O. Box 94, Grahamstown 6140, South Africa\\
$^{14}$ International Centre for Radio Astronomy Research, Curtin University, Perth, Australia\\
$^{15}$ National Radio Astronomy Observatory, Socorro, NM 87801, USA\\
$^{16}$ School of Physics and Astronomy, University of Manchester, Oxford Road, Manchester, M13 9PL, UK\\
$^{17}$ Physics Department, Lancaster University, Lancaster, LA1 4YB, UK\\
$^{18}$ Institute for Particle Physics and Astrophysics, ETH Z\"urich, Wolfgang-Pauli-Strasse 27, CH-8093, Z\"urich, Switzerland\\
$^{19}$ Modulos AG, Technoparkstr. 1, 8005 Zurich, Switzerland\\
}

\date{Released 2024 July 01}

\pagerange{\pageref{firstpage}--\pageref{lastpage}} \pubyear{2024}

\def\LaTeX{L\kern-.36em\raise.3ex\hbox{a}\kern-.15em
    T\kern-.1667em\lower.7ex\hbox{E}\kern-.125emX}

\begin{document}

\label{firstpage}

\maketitle

\begin{abstract}
Radio galaxies can extend far beyond the stellar component of their originating host galaxies, and their radio emission can consist of multiple discrete components. Furthermore, the apparent source structure will depend on survey sensitivity, resolution and the observing frequency. Associated discrete radio components and their originating host galaxy are typically identified through a visual comparison of radio and mid-infrared survey images. We present the first data release of Radio Galaxy Zoo, an online citizen science project that enlists the help of citizen scientists to cross-match extended radio sources from the Faint Images of the Radio Sky at Twenty Centimeters (FIRST) and the Australia Telescope Large Area Survey (ATLAS) surveys, often with complex structure, to host galaxies in 3.6~$\mu$m infrared images from the Wide-field Infrared Survey Explorer (WISE) and the Spitzer Space Telescope. This first data release consists of 100,185 classifications for 99,146 radio sources from the FIRST survey and 583 radio sources from the ATLAS survey. We include two tables for each of the FIRST and ATLAS surveys: 1) the identification of all components making up each radio source; and 2) the cross-matched host galaxies. These classifications have an average reliability of 0.83 based on the weighted consensus levels of our citizen scientists. The reliability of the DR1 catalogue has been further demonstrated through several parallel studies which used the pre-release versions of this catalogue to train and prototype machine learning-based classifiers. We also include a brief description of the radio source populations catalogued by RGZ DR1.

\end{abstract}


\section{Introduction}
Radio galaxies, a subset of Active Galactic Nuclei (AGN), may be single compact structures, simple doubles, or large, multi-component structures with complex morphologies.  Identifying which discrete radio components belong to a single source, and which optical or infrared galaxy is their host, can thus be a complex task.  For example, a line of three radio components might represent three individual AGN or star-forming galaxies, or a single radio AGN with a core and two extended radio lobes, or three possible combinations of a double radio source and an unrelated additional source.  A reliable characterisation of the radio source requires a recognition of both the underlying morphology and the host galaxy.

Morphological classifications of radio galaxies have traditionally been done by eye \citep[e.g.][]{norris06,Lin10,proctor11,andernach21,simonte22,simonte23}.  
Radio Galaxy Zoo\footnote{http://radio.galaxyzoo.org}(RGZ) is an online citizen science project that enlists the help of the public  to cross-match radio sources, often with complex structure, to host galaxies in infrared images \citep{banfield15}.  The online infrastructure and methodology of RGZ  is based on the original Galaxy Zoo project \citep{Lintott2008} which classified galaxy morphologies using optical observations from the Sloan Digital Sky Survey Data Release 7 \citep{york00,abazajian09}. The primary output from RGZ is the association of radio source components from an existing catalogue into single- or multiple-component radio sources as well as the identification of the host galaxy, if it is present.

The online classification part of RGZ concluded on 6 May 2019 after 5.5 years of operation.  In this period, over 12,000~registered users  contributed over 2.29~million radio source classifications for approximately 140,000 input ``subjects'' (as defined below).  In addition to providing classifications, interested RGZ citizen scientists had the option of using the RadioTalk\footnote{http://radiotalk.galaxyzoo.org/} forum, where they could post questions and ideas, and interact with the RGZ project scientists.  This forum led to serendipidous discoveries, such as the discovery of a new poor cluster of galaxies \citep{banfield16} and the identification of radio galaxies with  hybrid or giant radio galaxy morphologies \citep{kapinska17,tang20}.  The radio classifications enabled the investigation of cosmological alignments of radio sources \citep{contigiani17}, statistical studies of bent sources \citep{rodman18,garon19} and the exploration of machine learning-based methods for radio galaxy studies \citep[e.g.\ ][]{tang22,slijepcevic22,chen23}.  To date, a total of 15 papers have been based on this first generation RGZ project. Second generation RGZ projects such as Radio Galaxy Zoo LOFAR \citep{hardcastle23} and Radio Galaxy Zoo EMU \citep{bowles23} are also underway.

In this paper, we present the Radio Galaxy Zoo's Data Release 1 catalogue, which consists of 100,185 classifications for 99,146 radio sources from the  Faint
Images of the Radio Sky at Twenty Centimeters (FIRST) survey \citep{becker95} and for 583 radio sources from the  Australia Telescope Large Area Survey (ATLAS) survey \citep{norris06}.
This catalogue contains entries with well-defined parameters and reproducible quantities.  An expanded version of the catalogue, with additional entries that are limited by various biases, sample incompleteness and reliability uncertainties,  can be made available on request to the authors.

Section 2 defines the specific terminology used throughout this paper and briefly summarises the data samples. 
In Section 3, we describe the user classification tasks, and how the users' choices were used to build a consensus classification.  The DR1 catalogues are described in Section 4. Section~5 provides a brief description and demonstration of the source properties of the FIRST-WISE sample that are available from the RGZ DR1 catalogues.  We refer the reader to \citet{franzen15} and \citet{alger18} for a more detailed description of the ATLAS source properties.  Section 6 summarises this work.   Following \citet{banfield15}, we adopt a $\Lambda$ Cold Dark Matter cosmology of $\Omega_{\rm{m}}=0.3$, $\Omega_{\Lambda}=0.7$ and a Hubble constant of $H_{\rm{0}}=70$~km~s$^{-1}$~Mpc$^{-1}$ throughout this paper.  The WISE IR magnitudes refer to the Vega magnitude system.

\section{RGZ - Terminology \& sample}
This section defines the classification terminology and samples used for the RGZ project.  We also refer the reader to \citet{banfield15}, the RGZ project description paper that includes a more detailed discussion of the sample selection and project operations.

 
 \subsection{Terminology}
 \label{terminology}
 The following terminology is used throughout the paper, and where appropriate, was communicated to the users.
 \newline
  $\bullet$ Subject:  An image of a fixed size, 3$\arcmin$ {\bf{$\times$}} 3$\arcmin$ centered on an entry from the FIRST or ATLAS radio catalogue.  The subject is what is presented to the user, who classifies whatever is visible within the field.\\
  $\bullet$ Component:  A discrete patch of radio emission enclosed by a contour of constant brightness.\\
  $\bullet$ Peak:  A localised maximum within a component, identified only in the analysis stage. It should be noted that peaks need not be present in all components.\\
  $\bullet$ Source: A group of one or more components that the user considers to be part of the same physical system.\\
  $\bullet$ Host (counterpart): The infrared object selected by the users as the physical origin of the radio source.\\
  $\bullet$ Classification labels:  Sources are classified by their number of components and the total number of peaks within those components, in the analysis stage.  1c2p, for example would describe a source with one component (as defined above) with two peaks.  2c2p, by contrast, would be a two component source, each with one internal peak. In the case of a 2c2p, there could also be two components whereby one of the component consists of two peaks.  For the reasons of reproducibility, we do not follow traditional radio galaxy classes such as the Fanaroff-Riley (FR) types from \citet{fanaroff74}; but instead classify sources in terms of their observed radio structures, that is the number of associated components and peaks.  While it is possible to translate from the RGZ morphologies to FR classes, there is a large fraction for which clear FR classes can only be determined through further follow-up multiwavelength observations.   \\

\begin{figure}
\begin{center}
\includegraphics[width=3in]{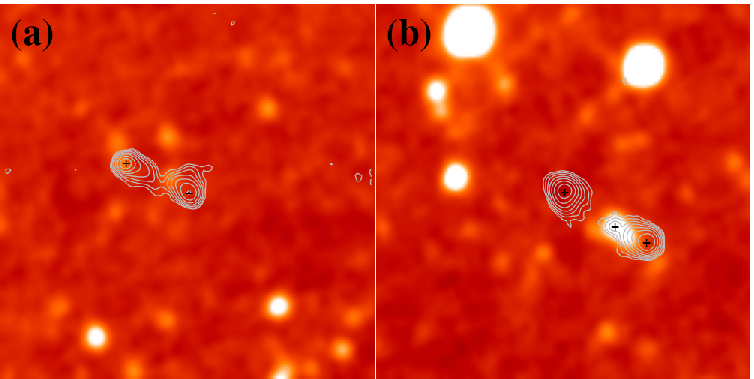}
\caption{RGZ DR1 morphologies are described in terms of the number of associated discrete radio components and flux density peaks (as described in Section~\ref{terminology}). Here we provide two examples: (a) a 1-component, 2-peaks source (1c2p); and (b) a 2-component, 3-peaks source (2c3p). The black crosshairs mark the locations of the peaks.}
\label{1c2p}
\end{center}
\end{figure}

 An illustration of these classifications is shown in  Figure~\ref{1c2p}.  On the left, a single set of contours enclose all the radio emission, so this source has one component.  Within the single set of contours, we find  two local maxima (peaks), so the source is classed as a source with 1-component and 2-peaks (1c2p).  On the right, by contrast, there is no single enclosing contour, so if these two components were classified by users as being physically associated, they would form a two component source.  One component has only a single localized peak, whereas the other has two peaks. Thus, the source would be labeled as 2c3p.
   
\subsection{The RGZ image sample}
RGZ is based primarily on position-matched radio image cutouts from the FIRST survey \citep[North of Declination~$+1.5$~degree; ][]{becker95, white97} and  the 3.4~$\mu$m ($W1$) AllWISE images from the Widefield Infrared Survey Explorer \citep[WISE; ][]{wright10,cutri13}.  A smaller additional sample (comprising 1.4\% of the total) was taken from the ATLAS Data Release 3 \citep{franzen15} and the 3.6~$\mu$m Spitzer Wide-area Infrared Extragalactic Survey \citep[SWIRE; ][]{lonsdale03}. In addition to our sample description in this section, we also refer the reader to the RGZ project description paper \citep{banfield15} for a more comprehensive description of the sample selection.

\subsubsection{The FIRST and WISE sample}
FIRST observed over 9000 square degrees of sky at a frequency of 1.4~GHz, a resolution of 5$\arcsec$, and a 1-$\sigma$ sensitivity of 150~$\mu$Jy~beam$^{-1}$, using the (Jansky) Very Large Array in New Mexico, USA \citep{white97}.  Objects from the survey, which we designate here as \emph{components},  were selected to be spatially resolved and have a sufficiently high signal-to-noise, resulting in an initial sample of approximately 200,000 components.   Based on noise fluctuations, approximately $15\%$ of these are expected to be unresolved \citep{banfield15}.  Such sources are useful for control purposes in an experiment such as RGZ, since they may also be classified (with host identifications)  using automated position-matching algorithms.

Using the online, and publicly-available dataset from the FIRST survey \citep{white97} that was available in the year before the launch of the RGZ project (December 2013), 3~arcminute by 3~arcminute cutouts were created and centred on each of the initial sample of approximately 200,000 components. The rms level for each of these cutouts was estimated and from this, contours were drawn from the lowest level of 4-$\sigma$ and increase by factors of 
$\sqrt{3}$. The radio images can be displayed either as a set of contours or as a blue colour scaled image.

The AllWISE $W1$ images have a resolution of 6.1$\arcsec$ \citep{cutri13}.  As shown in Figure~\ref{1c2p}, the AllWISE $W1$ images are near the confusion limit because the density of detected sources is larger than one in a few tens of beam areas.
This limits the reliability of host identifications to the brighter objects, as discussed in Section~3.3. In RGZ, the WISE $W1$ images were displayed with a log stretch using an orange (or `heat') colour scheme.

\subsubsection{The ATLAS and SWIRE sample} 
The ATLAS subjects are 2~arcminutes by 2~arcminutes in size, and the radio images from ATLAS have elliptical beamshapes that are either  $12.2\arcsec \times 7.6\arcsec$ or $16.8\arcsec \times 6.9\arcsec$ \citep{norris06,franzen15}. These ATLAS images are then cross-matched to 3.6~$\mu$m maps from SWIRE which have a resolution of $1.2\arcsec$ \citep{lonsdale03}.  

The justification for the smaller fields of view (FOV) for the ATLAS subjects comes from the design perspective of the project's online interface. As the SWIRE angular resolution is much finer than any of the other surveys, it was thought that the participants might have difficulty distinguishing between the unresolved IR sources for a larger FOV.  Furthermore, both the ATLAS and the SWIRE surveys are nearly a factor of ten more sensitive than the FIRST and WISE surveys, leading to greater visual confusion.  Similar to the FIRST sample, the ATLAS contours begin at 4-$\sigma$ and increase by factors of $\sqrt{3}$. An example of a subject based on the ATLAS and SWIRE sample is shown in Figure~\ref{egsubject}.

Even though the ATLAS subjects have smaller FOV than the FIRST subjects, there are in general a greater number of radio sources per ATLAS-based subject.  The median numbers of radio sources per subject for the ATLAS-based and FIRST-based subjects are three and one, respectively, noting that single component sources dominate the ATLAS sample. This suggests that source overcrowding and confusion may be  issues that affect the classification of ATLAS-based subjects, consistent with the results of \citet{alger18}. Further details about the ATLAS sample can also be found in \citet{franzen15} and \citet{alger18}.



\section{Classification, consensus and reliability}
In this section, we summarise the classification work performed by the users, and how their choices were used to establish a consensus classification for each source and a corresponding consensus value. 
Specific to this data release, we investigate the reliability of the user classifications as a function of consensus levels using expert verification on a limited sample.

\subsection{The User Classification Tasks} 
\begin{figure*}
\begin{center}
\includegraphics[scale=.62]{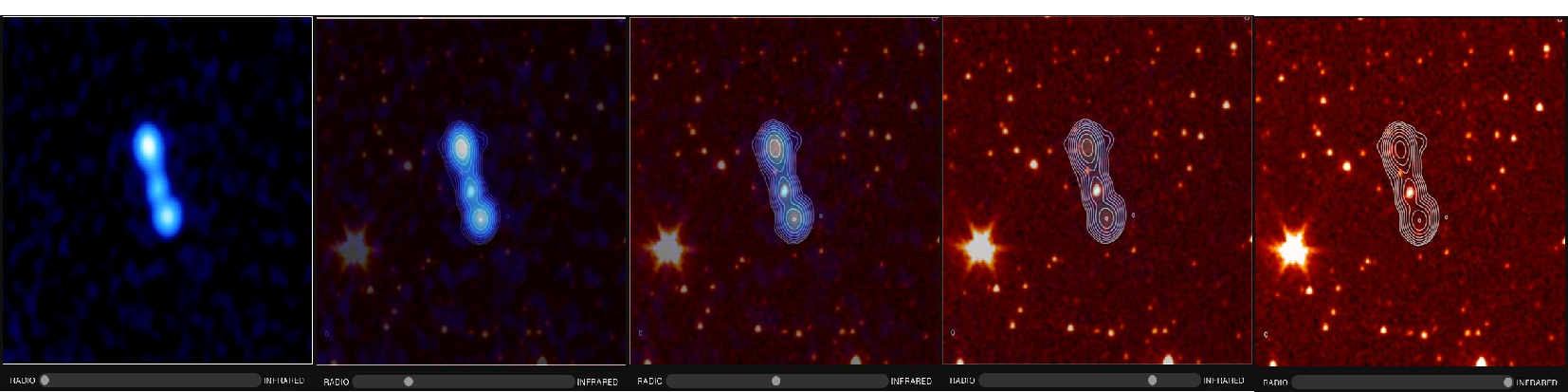}
\caption{Example of an ATLAS-SWIRE subject containing a single extended radio source as presented to the citizen scientists.  The five panels demonstrate the function of the slider tool to transition between the {\bf{$ATLAS$}} radio image (blue) and the {\bf{$SWIRE$}} infrared image (orange), on which the radio emission is marked as white contours. {\bf{The contours start at 4$\sigma$ and increase with factors of $\sqrt{3}$.}}}
\label{egsubject}
\end{center}
\end{figure*}

The RGZ user interface instructed users to examine the contoured radio component at the center of each subject.  In this process of examination,  users were encouraged to use the slider tool that is illustrated in Figure~\ref{egsubject}, to change the relative prominence of the radio (blue colour scale) and infrared (orange colour scale) emission. After this examination, the users are asked to carry out the following tasks:\\
$\bullet$ Identify (by clicking) any other radio components in the subject which appear to be associated with the central component;\\
$\bullet$ Click on the position of the likely infrared host galaxy associated with the central component and its associated radio components;\\
$\bullet$ Identify (by clicking) any significant radio components that appear to be other sources in the field (not associated with the central component), and all components associated with that additional source; \\
$\bullet$ Click on the position of the host of the additional source; \\
$\bullet$ When all additional sources and hosts have been identified, the user has the option to add comments, other diagnostic information, and raise questions about each source of interest through the ``Discuss" button which transfers them to the RadioTalk pages.  Alternatively, they can progress to the next subject.\\

Within the RadioTalk pages, users also had the option of looking at images that were twice as large in each dimension, and to examine the corresponding optical and higher sensitivity (but lower angular resolution) radio images from SDSS \citep{ahn14,alam15} and the NRAO VLA Sky Survey \citep[NVSS; ][]{condon98}  of the same field.  These additional images were used by a small number of users in the identification of the WAT described in \citet{banfield16}.  All of their ``clicks", however, are located (by design) within the original radio/IR subject.  

Figure~\ref{1c2p} provides an example of the classification choices that are faced by the users.  In Figure~\ref{1c2p}a, there is a clear double radio source, but since it is all enclosed within the lowest contour, a single click by the user would indicate that the entire area within the contour was chosen to indicate the full extent of the source.
In Figure~\ref{1c2p}b, by contrast, there are two isolated components, so the user would click on both of them, indicating that they were associated.  

In terms of identifying the infrared host position, for panel (a), users might click on the central infrared source located between the two radio lobes.  In panel (b), experienced users would recognize the blend of two or more infrared components near the center, and would likely click at the location of the radio core to indicate the likely host.  Less experienced users might click anywhere in the burnt-out infrared image overlapping with the radio contours.  Experts, and the most experienced users would check the corresponding SDSS image, and find that there is an optical galaxy coincident with the radio core, so come back to the original IR image and click at that location.

\subsection{Measuring Consensus}

Classification consensus refers to the agreement in source classification (the identification of associated radio components) between all users who have examined the same source. At the most simple level, the consensus level is the fraction of users who agreed upon a single source classification. In other words, if all users converged upon the same classification for the one source, that one source would have a consensus level of 1.0. Specifically for RGZ DR1, a user-weighted consensus level which takes into consideration the user weighting scheme (described in Section~3.2.2) is used to quantify the level of agreement or disagreement among the users.  \citet{banfield15} also provides a description of the data processing and for how source classifications are determined from the user identifications.


\subsubsection{Consensus algorithm}
In the simplest case where all users agree on the number of sources in the field, then the combination of components which is most commonly selected is chosen as the consensus classification.  The consensus level is then calculated as the ratio of the number of classifications which match the consensus to the total number of classifications.  The number of classifications used to estimate the consensus level also include user weightings (see Section~3.3). 

In the more common case where the users do not all agree on the number of sources in the subject, a more detailed process is followed.  For example, if there are three radio components in the subject, users could identify them as three separate sources, or one combined source, or three different permutations of two sources, one with a single component and one with two components.  In this case, the consensus algorithm first separates the classifications by the number of sources in the subject, $(N_\mathrm{s})$.  For each separate value of $N_{\mathrm{s},i}$, the most common assignment of components to the different sources is identified. The $N_{\mathrm{s},i}$ with the highest number of votes is deemed to be the consensus radio source.  Ties between vote counts are broken by randomly selecting among combinations with the same number of votes.

Once the consensus radio source(s) have been identified, the IR consensus (the agreement between the associated host to the source) is determined, based on the position in the subject indicated as the host position.  Data are only included for users whose radio classification for that source matches the consensus classification.   The  positions indicated by these users are used in a 2-D Gaussian kernel-density estimator (KDE) to estimate the probability density function of the host in pixel coordinates. If there are enough data to calculate the KDE (requiring at least 3 non-colinear points), we evaluate the KDE on the same grid size as the original infrared image and apply a $10\times10$~pixel maximum filter to locate peaks. There may be more than one peak in cases such as that shown in Figure~\ref{1c2p}, especially if multiple possible hosts are present.  The location of the  maximum of the probability distribution function is used for the position of the IR host.  
 WISE sources are identified as host galaxies when they are matched to within a 6-arcsecond radius of the RGZ IR host positions (see \citet{banfield15} for more details). 

If there are not enough data (fewer than three user classifications) to calculate the KDE then this algorithm will not be able to identify an IR host. Another disadvantage of the KDE method is that we are not able to identify more than one possible host galaxy for a radio source. Therefore, the RGZ DR1 catalogue contains fewer host galaxies in dense fields due to the inability for KDE to converge on a single locus of user clicks, when multiple host options are present.  
If the most common response for an individual radio source was to select ``No Infrared'', then the source is labeled as having no host.

\subsubsection{User weighting}
The above process is based on each user's classifications counting equally towards the final consensus for each source.  However, there is a broad range of experience among the users, which we sought to leverage. 
Among the registered users, the distribution of effort was highly unequal, with a Gini coefficient of $G=0.887$ \citep{glasser62}.  This indicates that the bulk of the source classifications are performed by the most prolific classifiers -- consistent with values measured for other citizen science projects \citep{cox15,spiers19} --  and signals the presence of a core group of dedicated users. 

We therefore made use of these different levels of experience by calibrating each registered user on their success in matching the classifications made by experts on a ``Gold Sample" (GS).  The GS is a set of 20 subjects which represent a range of classification complexities. These subjects are never withdrawn (see Section~3.2.3) and are presented at regular intervals to the users until all 20 have been completed by each individual user. As described in Section 4.2 of \citet{banfield15}, many RGZ participants will encounter several (if not all) of the Gold Sample (GS) sources from which their RGZ classifications can be calibrated and weighted. 

Anonymous users (not registered with the system) provided 25.4\% of the total classifications.  If a user is anonymous, their classifications cannot be collated.  If a user has classified fewer than five GS sources, and independent of whether these were correct or not, they are assigned one ``vote" in the weighted consensus. Users with sufficient GS classifications are then assigned a number of ``votes" from zero to five, based on how well they matched the expert classifications on the GS. For example, a user's classification will be upweighted to five votes if their GS classification is a perfect match to the experts' classification of the GS.

In practise, the vast majority of  RGZ classifications will be derived from weighted users because the majority of the classifications have been completed by the very experienced users.  In this subsection, we show the effect of user weighting on the FIRST sample. 
 
\begin{figure}
\includegraphics[scale=0.4]{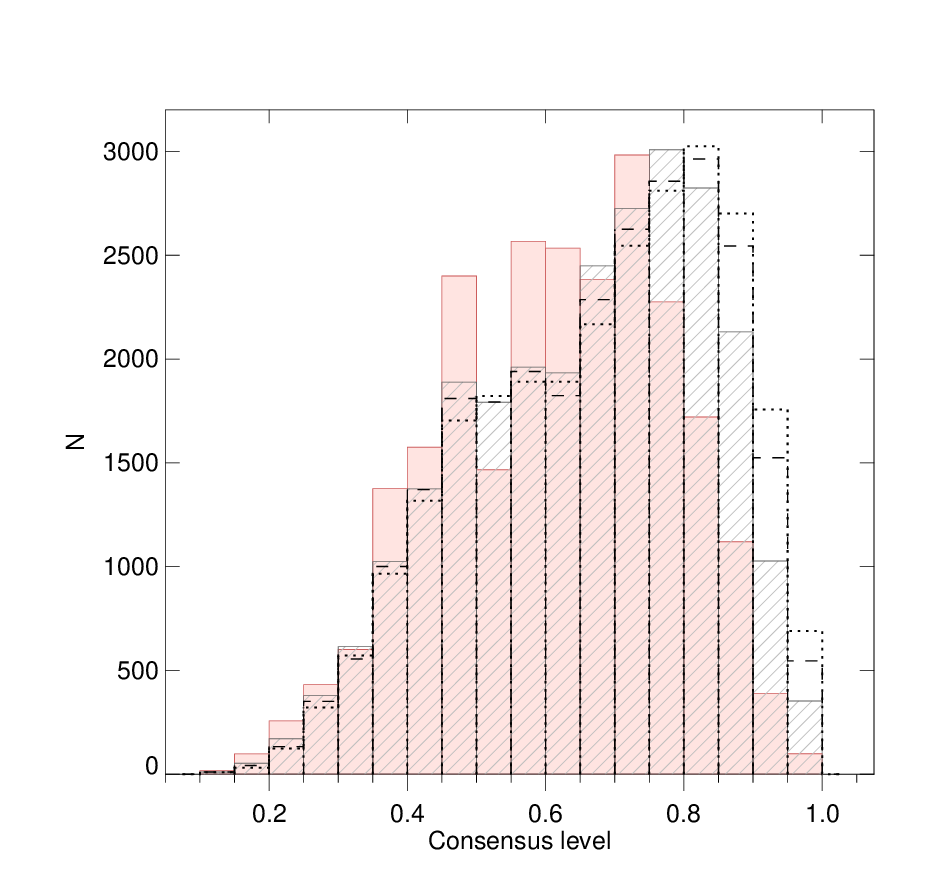}
\caption{Comparison of consensus level distributions for 2-component sources for catalogues with no user weighting (pink shaded histogram) with respect to catalogues which adopt three different user maximum weightings (+5, +10 and +20).  The catalogue that adopts the +5 weighting is represented by the distribution filled with diagonal lines.  The catalogues with +10 and +20 weightings are represented by the distributions outlined by dashed and dotted lines, respectively.  }
\label{ncompcatcomp}
\end{figure}

Prior to the application of user weighting, we find the raw mean and median consensus levels for single component sources to be 0.86 and 1.0, respectively. It is unsurprising that the classifications of simple single component sources result in excellent consensus from our users.   For RGZ sources with two components, the median consensus value increased for 0.63 to 0.69 when weights were applied. The shift in the distribution of consensus levels due to weights are shown in Figure~\ref{ncompcatcomp}, which also illustrates the results of higher levels of weighting (up to 10 and 20), with which we experimented. The higher level weightings led to little or no improvement for two component sources, so we adopted a maximum weighting of five. With this maximum weighting, the consensus levels for three to five components increased by similar amounts to the two component case.

The majority of the FIRST sample (99.9\%) have 5 or fewer radio components. In general, when the number of components and angular size increase, the consensus fraction decreases. The weighted catalogue results in an increase of consensus levels for multiple component sources. However, the maximum angular sizes (or largest angular extents) for the FIRST sample are not changed by the implementation of user weightings.  
  
Based on the improvements with the use of weights, the weighted consensus level is used, hereinafter and for DR1, and simply referred to as the ``consensus'' or ``consensus level''.

\subsubsection{Retirement}
Once an individual subject has enough user classifications to establish a consensus value, the object is ``retired," i.e., it is not shown to any more users.  This enables the efficient classification of subjects.   This does not mean that a high level of consensus has been reached;  a lack of agreement among an initial group of users would generally persist if more user classifications were obtained, and the consensus level would still be low.

The limit for retiring an individual subject was initially set at 20~classifications for every image \citep{banfield15}. However, based on early results, we found that there was little or no added value beyond 5~classifications for the simplest cases of a single component in a subject.  That 5-classification threshold was thus adopted for single components on 20 Jun 2014, after $\sim750,000$~classifications were completed. For more complex sources, a retirement threshold of 20 remained.

\subsection{Consensus thresholds and reliability for the FIRST sample}
The consensus levels are an indicator of the reliability of source classifications.  In this section, we examine the distribution of consensus levels and assess how the reliability depends on that level.  Choosing a threshold consensus level for inclusion in the catalogue inevitably involves compromises between the number of sources that can be studied and their reliability.  We therefore set a threshold consensus level for DR1 which we believe to provide sufficiently large and reliable samples for many science studies.  It is possible to also examine classifications with consensus below the DR1 threshold; this is beyond the scope of this paper.

\begin{figure}
\begin{tabular}{c}
\includegraphics[scale=0.39]{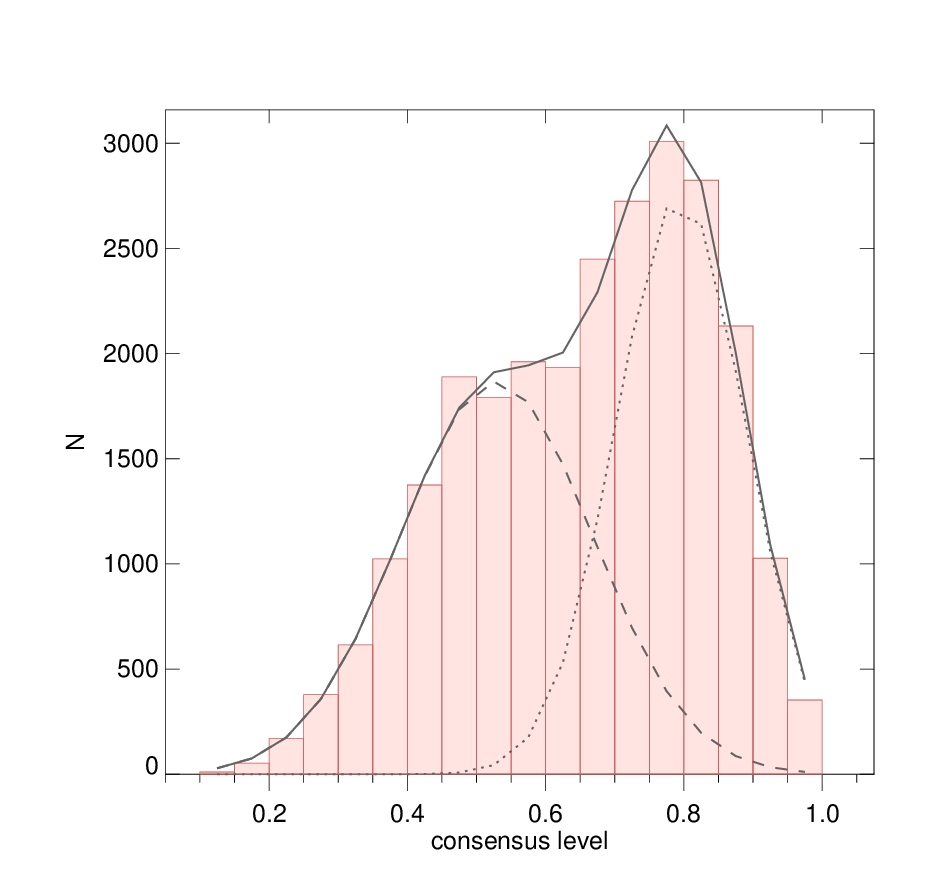}\\
\includegraphics[scale=0.4]{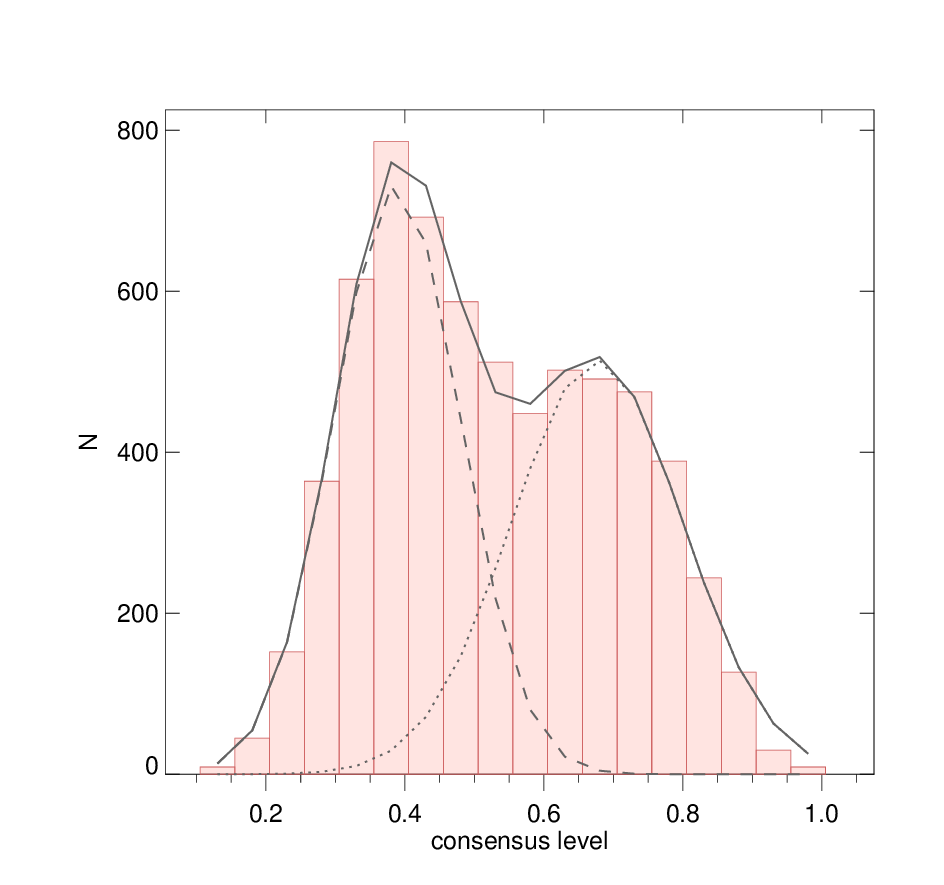}\\
\end{tabular}
\caption{Distribution of weighted consensus level for two component sources (top panel) and three component sources (bottom panel) in the FIRST sample.  The low and high consensus populations are represented by the dashed and dotted distributions, and the distribution outlined by the thick grey solid line is the sum of the two distributions.}
\label{doubclwfit}
\end{figure}

The distributions of (weighted) consensus levels for all 2-component and 3-component sources are shown in Figure~\ref{doubclwfit}.  These are broad, asymmetric distributions which we model as the sum of two populations for each class:  low level and  high  level consensus distributions.  There is no natural consensus level at which a distinct high level population begins to dominate, so independent of the detailed modeling, there is substantial overlap in levels associated with the two populations.  For both two and three component sources, the two populations are better separated with the use of user weighting.

\subsubsection{Reliability using expert classifications}

\begin{figure}
\includegraphics[scale=0.2]{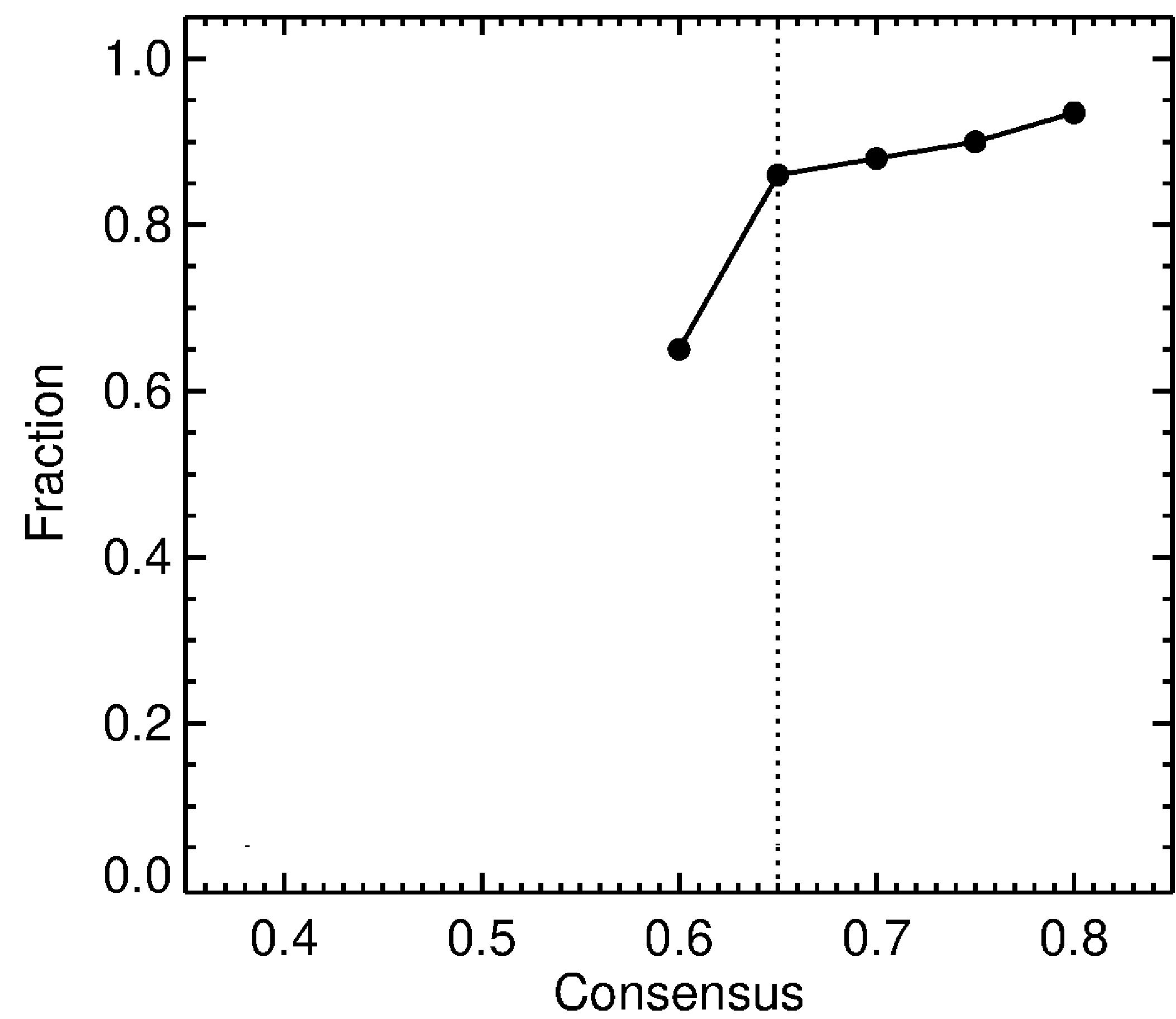}
\caption{The fraction of classifications (from the test sample of 1000) 
that are verified to be correct by the science team as a function of 
catalogued consensus levels.  The vertical dotted line shows the
threshold of consensus required for inclusion into this data release. }
\label{ivycheck}
\end{figure}

To characterise the reliability of the DR1 sources as a function of consensus level, we assessed the agreement between the radio morphology consensus classifications and the corresponding classifications provided by experts.  We selected 
a random sample of 1000 DR1 sources with user consensus levels between 0.6 and 0.8, in the overlap region between the two assumed populations where we had the most uncertainty about the reliabilities. These included one-, two- and three-component sources.  
A subset of the science team (OIW, LR, HA, SS \& RN; hereafter known as verifiers) then independently classified the 1000 sources, and an expert radio morphology consensus classification was determined for each source. 
These verifiers were shown a single IR image overlaid with the FIRST radio contours and were asked to verify if the DR1 classification is plausible or not. The verifiers used the IR information to help classify the radio morphology, but did not separately indicate the corresponding IR host.  Disagreement in source classifications occured between the verifiers for approximately 10~percent of the sources --- indicating that the level of irreducible subjectivity in these measures.
Comparing the user and verifier consensus classifications, we found them to be in agreement $\approx$85\% of the time when the user consensus level was above 0.65.  As can be seen in Figure~\ref{ivycheck}, the agreement between users and experts significantly drops below a level of 0.65, as expected from the distributions in Figure~\ref{doubclwfit}. Hence, we therefore establish a consensus level threshold of 0.65 for RGZ DR1. We note that earlier analysis by \citet{banfield15} followed the principles of Galaxy Zoo's `clean' sample consensus cutoff of 0.8 \citep{Lintott2008,willett13} and thus used a more conservative level of 0.75.  

Using a consensus level of 0.65 as our consensus threshold, DR1 has successfully enabled early efforts to utilize machine learning experiments for automated radio source classification \citep{lukic18,alger18}.  For example, the reliability of RGZ DR1 is consistent with the mean average precision obtained for ClaRAN \citep[a deep-learning based radio morphology prototype end-to-end source classifier; ][]{wu19}. As a result, recent reviews \citep{becker21, huertas22} for the application of machine learning methods to astronomy have commended the use of citizen science catalogues such as RGZ DR1 as valuable training datasets for training supervised learning models.

\subsubsection{Host galaxy effect on reliability}
\begin{figure}
\includegraphics[scale=.6]{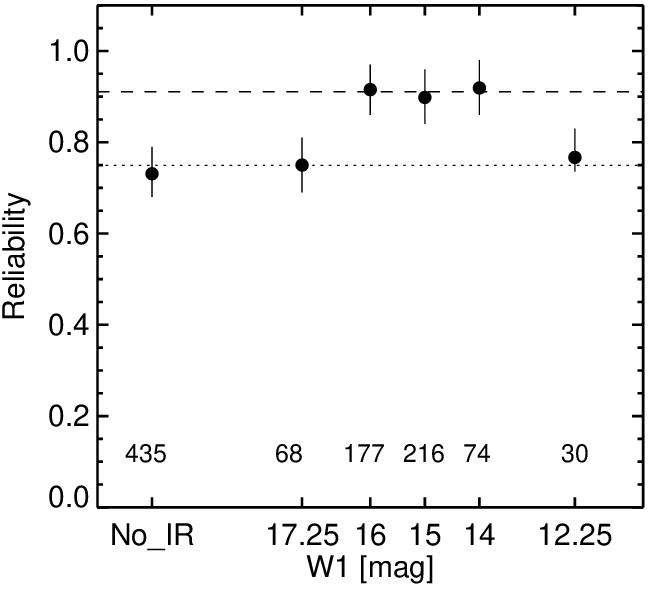}
\caption{Reliability as a function of the WISE W1 magnitude, averaged over all radio classifications.  The average reliability above and below magnitude 17 are indicated by horizontal lines. We also show the number of sources per bin at the bottom of the plot. It should be noted that 'No\_IR' represents the sources for which no IR host has been identified.}
\label{rew}
\end{figure}

The reliability of the radio classification may also depend on the the relative distribution of potential host galaxies traced by the infrared image. Here  we investigate whether the reliability of the radio source classifications was dependent on the presence of a bright infrared host.  Hosts, as described above, were identified by the peak in the KDE distribution of user clicks identifying the host position.   For our sample of 1000 expert classified sources, approximately half were identified with hosts brighter than W1 magnitude 17, and half either fainter or having no consensus host.   We set up bins in W1 magnitude, and  performed a weighted average over all the various radio classifications and their respective reliabilities for hosts in that bin.  This produced a characteristic radio reliability as a function of $W1$ magnitude\footnote{Reliability, $R(W1) = \frac{\sum_{cl=0}^{1}  ( N(cl,W1) \times F(cl,W1)) }{ [\sum_{cl=0}^{1} (N(cl,W1))]}$, where $F$ is the verified fraction, $N$ is the number of sources (Figure~\ref{ivycheck}), $cl$ is the weighted consensus level and $W1$ is the WISE $W1$ magnitude.} as shown in Figure~\ref{rew}. 

The mean classification reliability for sources with hosts brighter than $W1$ magnitude of 17 but fainter than 13 was  $0.91 \pm 0.01$, while the mean reliability for fainter or no infrared hosts (or brighter than $W1$ magnitude of 13) was $0.75 \pm 0.02$.  This means that the DR1 sample has a mean reliability of 0.83. Thus, the presence of a bright host did increase the reliability somewhat, but was not essential for a reliable radio classification. In fact, $W1$ hosts brighter than 13 magnitudes result in reduced reliability, possibly due to source saturation or confusion in the reference IR image.  It is likely that a more careful investigation of this dependence with large samples of each type of radio source would find niches where bright hosts made a significant difference.

\subsection{Consensus thresholds and reliability for the ATLAS sample}
\begin{figure}
\includegraphics[scale=0.5]{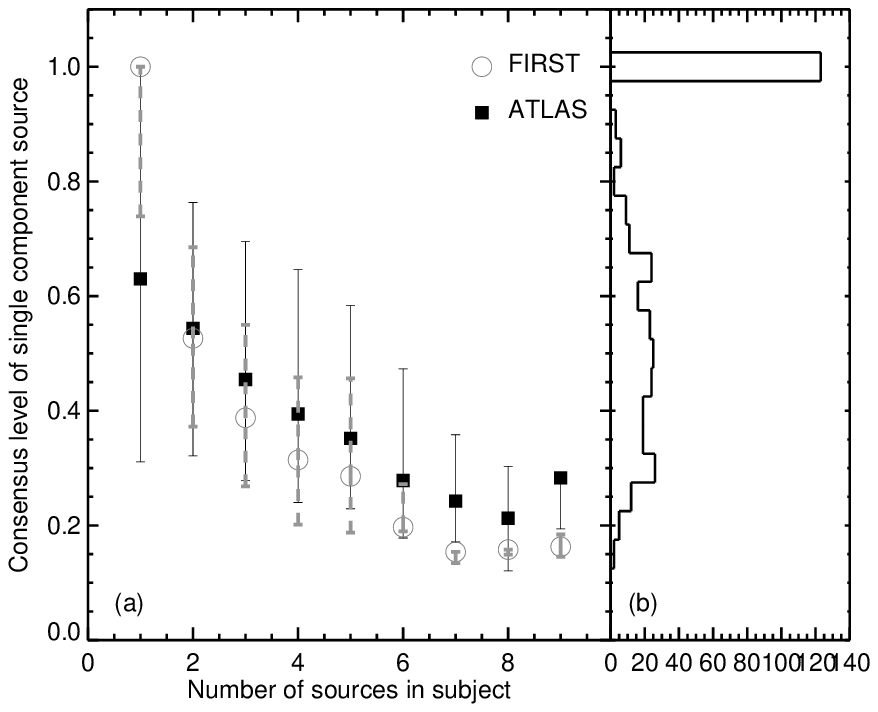} \\
\caption{Panel (a): Median consensus levels for single component radio sources in the FIRST-based (grey open circles) and the ATLAS-based (solid black squares) subjects. The thick grey dashed error bars and the thin black error bars show the 10th-and 90th-percentile consensus levels for the FIRST-based  and the ATLAS-based subjects, respectively. Panel (b): Distribution of consensus levels for isolated single component ATLAS sources. }
\label{clvsn}
\end{figure}

The main impact of user weighting on the  RGZ classifications of the ATLAS sources is the upward-shift in consensus levels after the weighting is introduced --- comparable to the effect of user weighting for the FIRST sample.  
To explore the impact of multiple sources and confusion, we examine the weighted consensus levels of isolated single-component radio sources (ie. single component radio sources which happen to be the only radio source within the subject). Such a source is likely to be the simplest form of  classification for the participant.  Figure~\ref{clvsn}a shows the median consensus levels for single component sources as a function of the number of radio sources within the same subject. The general result here is that the median consensus levels decreases with increasing number of sources per subject.  As can be seen from Figure~\ref{clvsn}a, there exists a larger scatter (as shown by the errorbars which indicate the 10th- and 90th- percentile levels) in consensus levels for the ATLAS-based sample relative to that of the FIRST sample.  

The median weighted consensus level for an isolated single-component ATLAS-based radio source is only 0.63 (relative to 1.0 for the FIRST sample). The consensus level distribution  for isolated single-component ATLAS-based radio sources shows that the number of isolated single-component ATLAS-based radio sources with consensus levels less than 0.8 is a factor of 1.6 greater than the number of isolated single-component ATLAS-based radio sources with consensus levels above 0.8 (Figure~\ref{clvsn}b).



With the exception of one two-component radio source, the remaining ATLAS-DR1 sources are all single-component compact radio sources.  The large number of ATLAS-based
compact radio sources is due to the initial RGZ sample selection which included a random sample
of ATLAS sources that is not biased towards more extended source morphologies \citep[unlike
 the selection made for the FIRST-based sample; ][]{banfield15}. The percentage of extended sources for the ATLAS sample is approximately 17~percent \citep{franzen15}.

\begin{figure}
\includegraphics[scale=0.6]{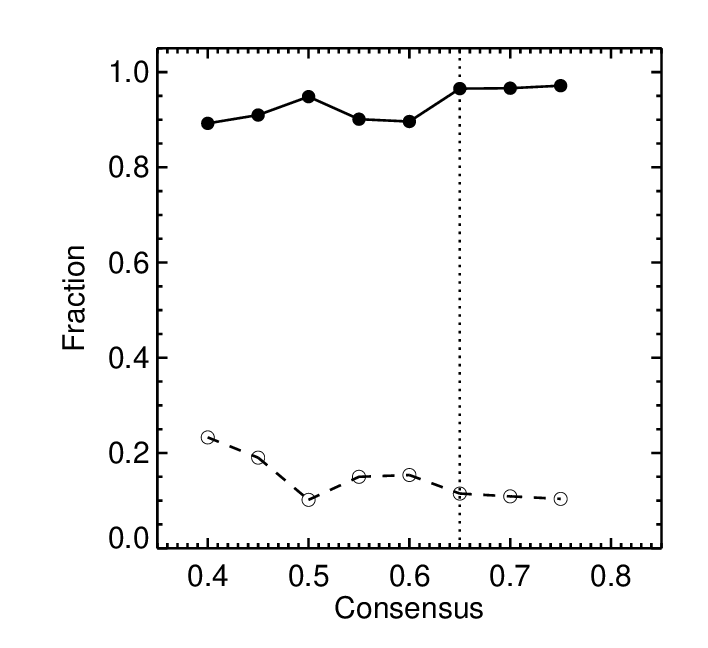} \\
\caption{Verified fraction (solid line) and the cross-match confusion fraction (dashed line) of ATLAS subjects as a function of consensus levels. Cross-match confusion can arise when multiple IR sources are coincident with, and could be matched to a single ATLAS radio source.}
\label{ivyatlascheck}
\end{figure}

What is driving the low consensus levels for single component sources in the ATLAS sample? Figure~\ref{atlaslowcl} in Appendix~\ref{atlasappendix} presents examples of isolated single-component ATLAS souces with the 12 lowest consensus levels.  Visual inspection of the isolated single-component ATLAS-based radio sources with low consensus levels (Figure~\ref{atlaslowcl}) suggests that the uncertainty in classifications may be due to the disparity in angular resolution of the radio and IR images that are being cross-matched.  In other words, it is possible to match the radio emission with multiple host galaxies.  Furthermore the greater uncertainty in the centre position of the radio emission relative to the IR resolution is likely to result in a perceived misalignment between the position of the radio emission and that of the corresponding host galaxy.  

To verify the reliability of the DR1 classifications for the ATLAS sample, a subset of the ATLAS-DR1 subjects that are based upon the ATLAS  survey were independently examined by the science team (OIW \& JS). We find that the fraction of agreement in classifications between the science team and DR1 is uniformly high (agreement fraction of 0.9 and greater) regardless of DR1 consensus levels (Figure~\ref{ivyatlascheck}).  We attribute the high agreement (or consensus) fractions to the dominance of single-component compact sources within the ATLAS-DR1 sample.  Due to the larger difference in angular shapes and resolution between the ATLAS and SWIRE observations, there may be multiple IR sources which can be matched to the position of the ATLAS radio source.  We refer to this as a cross-match confusion.  
 In Figure~\ref{ivyatlascheck}, we show the fraction of sources which are affected by cross-match confusion (where it is possible to match multiple IR hosts to a single radio source), as open circles (connected by the dashed line).  
 Similar to the recovered fraction of classification agreement, the fraction of confused ATLAS
test subjects is relatively flat at a 10~percent level at consensus levels of greater than
 or equal to 0.65, and increases to approximately 23~percent at a consensus level of 0.4.
While confusion affects a significant fraction of the ATLAS-SWIRE cross-identifications, the
fraction in agreement is high because the majority of the ATLAS-DR1 sources are
single-component sources. Therefore the ATLAS-DR1 sources represent the control
sample of classifications for the Radio Galaxy Zoo project, rather than the type
of radio sources that benefits most from visual identification.

\section{RGZ Data Release 1}
The Radio Galaxy Zoo Data Release 1 consists of 100,185  radio source classifications (from both the the FIRST and ATLAS samples) with user weighted consensus levels greater or equal to 0.65; this represents approximately 70~percent of the full sample of classifications that were derived from the project.  

In RGZ DR1 catalogues, there are more classifications than radio sources because: 1) the same extended radio source may appear in more than one subject; and 2) there are multiple possible source classifications that share the same radio component  \citep[e.g.\ Class C sources described in ][]{banfield15}. Classifications that relate to the same radio sources are cross-associated and flagged within the radio component catalogue.

The main data products that we provide in this data release are: 1) the radio source `morphology', described by the number of components
and peaks each source comprises (Table~\ref{dr1morphtab_first} and  Table~\ref{dr1morphtab_atlas}); and 2) the possible IR host counterpart (Table~\ref{dr1hosttab_FIRST_1}, Table~\ref{dr1hosttab_FIRST_2}, Table~\ref{dr1hosttab_ATLAS_1} and Table~\ref{dr1hosttab_ATLAS_2}).  
As described in Section~2.1, our catalogue defines radio source morphology in terms of the number of associated radio source components and peaks in emission (as observed from the FIRST or ATLAS images).

We present the measured consensus levels for each classification generated by the project as opposed to devising an automated method to estimate the combined consensus level from multiple classifications for each radio source.  For example, multiple sources may share individual components in different classifications. Due to the irreducible ambiguity in the dataset, it may not be possible to determine which source is truly associated with any one component via our current use of the KDE method. However it is reassuring that we are able to report the statistical reliability of this catalogue that is empirically demonstrated by earlier RGZ publications.

We describe the RGZ FIRST radio morphology and host galaxy catalogues in Sections~4.1 and 4.2, respectively.   

\subsection{FIRST-WISE radio source morphology catalogue}
\begin{figure}
\includegraphics[scale=0.5]{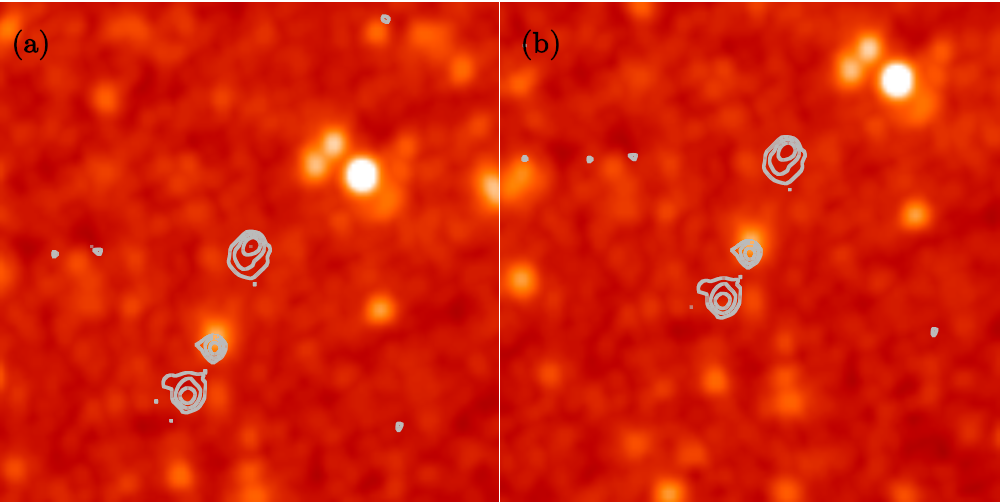} \\
\caption{An example pair of Zooniverse subjects which have resulted in a pair of duplicate classifications within RGZ DR1. The radio emission is overlaid as grey contours over the WISE $W1$ image in both panels. The contours start at 4$\sigma$ and increase with factors of $\sqrt{3}$. Panels (a) and (b) show the Zooniverse subjects ARG00025jk and ARG00025kh, respectively. The host galaxy for these subjects is the same, WISEA~J103944.16+231303.1.}
\label{egdup}
\end{figure}

The RGZ DR1 radio source morphology catalogue for the FIRST sample lists 16 parameters pertaining to each radio source classification. Table~\ref{dr1morphtab_first} presents the first 10 lines of the FIRST-based radio morphology catalogue.  The full machine-readable version of this table is included in the supplementary material of this paper (filename: `DR1\_FIRST\_radio\_classifications.csv').  The 16 columns presented are:  
\newline {\em{Column 1}}: the catalogue identification number or index
\newline {\em{Column 2}}: the Radio Galaxy Zoo source identifier in the IAU-accepted format of RGZ Jhhmmss.s+ddmmss, determined by either: 1) the WISE counterpart; or in its absence, 2) the centre position of a box that encapsulates the entire radio source where a WISE counterpart has not been identified. It should be noted that the RGZ names are truncations (as opposed to rounding) of the sexagesimal positions.
\newline {\em{Column 3}}: the Zooniverse subject identifier, corresponding to the subject identity from the web interface of the project\footnote{Each Zooniverse subject can be found online via the URL {\tt{https://radiotalk.galaxyzoo.org/\#/subjects/ARG*******}} where the asterisks represent the individual identifier characters.}
\newline {\em{Columns 4 \& 5}}: right ascension (J2000) and declination (J2000) of the radio source centre (which does not necessarily correspond to the host position)
\newline {\em{Column 6}}: number of votes for this source classification (includes user weighting)
\newline {\em{Column 7}}: total number of votes relevant to this source (includes user weighting)
\newline {\em{Column 8}}: consensus level (CL) for this radio source classification (includes user weighting)
\newline {\em{Column 9}}: number of distinct radio source components
\newline {\em{Column 10}}: total number of radio source peaks. To estimate the total number of peaks, a local maximum filter and a binary erosion function from {\tt{SciPy}} \citep{virtanen20}  was used to extract the location of radio intensity peaks within a source. The sum of these peaks provide the total number of peaks within a source.
\newline {\em{Column 11}}: largest angular extent (LAE) of the radio source in arcseconds,   the diagonal of a rectangle that encompasses the entire radio source at the level of the lowest (4$\sigma$) radio brightness contour. 
\newline {\em{Column 12}}: total solid angle (TSA) of the radio source in arcseconds squared,  the sum of the area contained within the 4$\sigma$ contour level
\newline {\em{Column 13}}: outermost level (OL) radio brightness in mJy~beam$^{-1}$, corresponding to the lowest contour level which is  4-sigma above the noise of the radio image
\newline {\em{Column 14}}: total flux (TF) of the radio source in mJy,  estimated from the sum of the flux in all associated source components
\newline {\em{Column 15}}: total flux error (TF~err) or uncertainty in TF, in mJy, calculated from the summed quadrature of the flux uncertainties for each of the associated components
\newline {\em{Column 16}}: Duplicate Components (DC), which identifies the other catalogue index (or indices) with which this radio source has component(s) in common.  Duplicate classifications can occur as a radio source may be associated with more than one entry in the FIRST catalogue \citep{white97}.  See Figure~\ref{egdup} for an example of a pair of duplicate classifications from two Zooniverse subjects, ARG00025jk (panel a) and ARG00025kh (panel b).   ARG00025jk and ARG00025kh relate to the same source and the same host, but are centred on  different radio source components. 

\subsection{FIRST-WISE host galaxy catalogue}
The FIRST-based DR1 host galaxy catalogue tabulates 23 parameters pertaining to the host galaxy that has been cross-matched to the classified radio sources.  Table~\ref{dr1hosttab_FIRST_1} and Table~\ref{dr1hosttab_FIRST_2} present these 23 parameters for the first 10 lines of the host catalogue.  The full machine-readable version of Table~\ref{dr1hosttab_FIRST_1} and Table~\ref{dr1hosttab_FIRST_2} are included in the supplementary material of this paper (filename: `DR1\_FIRST\_host\_properties.csv').    The first three columns of Table~\ref{dr1hosttab_FIRST_1} are the same as those of Table~\ref{dr1morphtab_first}.  We also note that WISE band magnitudes without uncertainties are to be understood as upper limit values.   A description of WISE infrared host properties provided for the FIRST-based classifications are as follows:
\newline {\em{Column 1}}: the catalogue identification number or index
\newline {\em{Column 2}}: the Radio Galaxy Zoo source identifier
\newline {\em{Column 3}}: the Zooniverse subject identifier. This identity corresponds to the subject identity from the web interface of the project
\newline {\em{Column 4}}: right ascension in degrees (J2000) of the host location via participant consensus 
\newline {\em{Column 5}}: declination in degrees (J2000) of the host location via participant consensus 
\newline {\em{Column 6}}: the AllWISE infrared source identifier of the host galaxy
\newline {\em{Column 7}}: the AllWISE source right ascension (J2000) in degrees
\newline {\em{Column 8}}: the AllWISE source declination (J2000) in degrees
\newline {\em{Column 9}}: 3.4~$\mu$m WISE Band 1 ($W1$) magnitude 
\newline {\em{Column 10}}: $W1$ magnitude uncertainty
\newline {\em{Column 11}}: $W1$ signal-to-noise ratio 
\newline {\em{Column 12}}: 4.6~$\mu$m WISE Band 2 ($W2$) magnitude 
\newline {\em{Column 13}}: $W2$ magnitude uncertainty
\newline {\em{Column 14}}: $W2$ signal-to-noise ratio 
\newline {\em{Column 15}}: 12~$\mu$m WISE Band 3 ($W3$) magnitude 
\newline {\em{Column 16}}: $W3$ magnitude uncertainty
\newline {\em{Column 17}}: $W3$ signal-to-noise ratio 
\newline {\em{Column 18}}: 22~$\mu$m WISE Band 4 ($W4$) magnitude 
\newline {\em{Column 19}}: $W4$ magnitude uncertainty
\newline {\em{Column 20}}: $W4$ signal-to-noise ratio 
\newline {\em{Column 21}}: number of matches to the AllWISE catalogue ($N^{\rm{WISE}}_{\rm{MATCH}}$)
\newline {\em{Column 22}}: Duplicate WISE Match (DWM), which identifies the catalogue index of any other catalogued classification(s) which share the same WISE host.  This duplication arises when the same radio source and host appears in more than one subject
\newline {\em{Column 23}}: Photometric redshifts ($z_{\rm{phot0}}$) of the matched WISE host from \citet{beck22}   

\subsection{RGZ DR1 ATLAS catalogues}
The DR1 ATLAS radio morphology catalogue for the ATLAS sample consists of the same 16 parameters that describe the DR1 FIRST sample (see Section~4.1).  Table~\ref{dr1morphtab_atlas} presents the first 10 rows of the DR1 radio morphology catalogue for the ATLAS sample.

The infrared host properties of the ATLAS-based DR1 sources originate from the SWIRE survey \citep{lonsdale03}.  The infrared photometry provided as part of this data release is derived from the SWIRE version 2 and 3 catalogues \citep{surace05}. Table~\ref{dr1hosttab_ATLAS_1}  and Table~\ref{dr1hosttab_ATLAS_2} present the information for the first 11 columns, and that for columns 12 to 19, for the first 10 lines of this table respectively.  The full machine-readable version of this table is included in the supplementary material of this paper (filename: `DR1\_ATLAS\_host\_properties.csv'). The 19 host properties and parameters provided by this host catalogue are as follows:
\newline {\em{Column 1}}: the catalogue identification number or index
\newline {\em{Column 2}}: the Radio Galaxy Zoo source identity
\newline {\em{Column 3}}: the Zooniverse subject identifier. This identity corresponds to the subject identity from the web interface of the project
\newline {\em{Column 4}}: Right ascension in degrees (J2000) of the host location via participant consensus 
\newline {\em{Column 5}}: Declination in degrees (J2000) of the host location via participant consensus 
\newline {\em{Column 6}}: the SWIRE source identifier of the host galaxy
\newline {\em{Column 7}}: the SWIRE source right ascension (J2000) in degrees
\newline {\em{Column 8}}: the SWIRE source declination (J2000) in degrees
\newline {\em{Column 9}}: aperture-corrected 3.6~$\mu$m Spitzer IRAC Band 1 ($f_{\rm{3.6}}$) flux in $\mu$Jy. The flux is measured from an aperture with a radius of 1.9\arcsec.
\newline {\em{Column 10}}: $f_{\rm{3.6}}$ flux uncertainty in $\mu$Jy
\newline {\em{Column 11}}: aperture-corrected 4.5~$\mu$m Spitzer IRAC Band 2 ($f_{\rm{4.5}}$) flux in $\mu$Jy. The flux is measured from an aperture with a radius of 1.9\arcsec.  
\newline {\em{Column 12}}: $f_{\rm{4.5}}$ flux uncertainty in $\mu$Jy
\newline {\em{Column 13}}: aperture-corrected 5.8~$\mu$m Spitzer IRAC Band 3 ($f_{\rm{5.8}}$) flux in $\mu$Jy. The flux is measured from an aperture with a radius of 1.9\arcsec.  
\newline {\em{Column 14}}: $f_{\rm{5.8}}$ flux uncertainty in $\mu$Jy
\newline {\em{Column 15}}: aperture-corrected 8.0~$\mu$m Spitzer IRAC Band 4 ($f_{\rm{8.0}}$) flux in $\mu$Jy. The flux is measured from an aperture with a radius of 1.9\arcsec.  
\newline {\em{Column 16}}: $f_{\rm{8.0}}$ flux uncertainty in $\mu$Jy
\newline {\em{Column 17}}: Number of matches to the SWIRE catalogue ($N^{\rm{SWIRE}}_{\rm{MATCH}}$)
\newline {\em{Column 18}}: spectroscopic redshift ($z_{\rm{sp}}$) of the matched SWIRE host where available \citep{rowanrobinson13}
\newline {\em{Column 19}}: photometric redshift ($z_{\rm{ph}}$) of the matched SWIRE host where $A_{\rm{V}}$ is modelled as a free parameter \citep{rowanrobinson13}

\section{DR1 properties}
RGZ DR1 is one of the largest catalogues of extended radio sources ever compiled through visual classification. In this section, we focus on the properties of the FIRST-based catalogues, which accounts for  99.4\% of DR1. Comparisons to previously known correlations further verifies the scientific readiness of this data release.

\clearpage

\begin{landscape}
{\bf{
\begin{table}
\caption{The first 10 lines of the FIRST-based radio source morphology classification catalogue.}
\label{dr1morphtab_first}
\begin{tabular}{cccccccccccccccc}
\hline
Cat ID & RGZ ID & Zooniverse ID & RA & Declination & $N_{\rm{votes}}$ &$N_{\rm{total}}$ & CL & $N_{\rm{comp}}$ & $N_{\rm{peaks}}$ & LAE & TSA & OL & TF & TF err & DC  \\
            &         &              &  deg & deg &             &              &    &               &               & arcsec & arcsec &mJy~beam$^{-1}$& mJy & mJy & \\
(1) & (2) & (3) & (4) & (5) & (6) & (7) & (8) & (9) & (10) & (11) & (12) & (13) & (14) & (15) & (16)  \\
\hline
5  &RGZ\_J143121.3+232251 &ARG0002565 &217.83890 &23.38094 &38 &38 &1.0 &1 &1 &18.6 &128.56 &0.81  &29.79 &0.19 &\ldots\    \\
6  &RGZ\_J140600.1+232249 &ARG0002566 &211.50053 &23.38030 &39 &39 &1.0 &1 &1 &13.9 &69.95  &0.72  &3.97  &0.12 &\ldots\       \\
7  &RGZ\_J105512.0+232306 &ARG000255x &163.79966 &23.38497 &57 &57 &1.0 &1 &1 &37.4 &304.39 &0.89  &18.59 &0.32 &\ldots\      \\
8  &RGZ\_J084831.6+232245 &ARG0002567 &132.13167 &23.37924 &32 &32 &1.0 &1 &1 &15.2 &88.86  &0.73  &7.14  &0.14 &\ldots\         \\
9  &RGZ\_J171515.3+232247 &ARG0002568 &258.81516 &23.37785 &38 &38 &1.0 &1 &1 &27.4 &151.25 &0.77  &7.77  &0.2  &\ldots\       \\
10 &RGZ\_J164643.0+232254 &ARG000255v &251.67924 &23.38211 &41 &42 &0.98&1 &1 &16.6 &105.87 &0.93  &10.37 &0.2  &\ldots\    \\
11 &RGZ\_J091550.4+232254 &ARG000255y &138.96043 &23.38164 &47 &47 &1.0 &1 &1 &20.1 &141.8  &0.88  &15.08 &0.22 &\ldots\    \\
15 &RGZ\_J095705.6+232253 &ARG0002560 &149.27362 &23.38166 &40 &40 &1.0 &1 &1 &12.3 &56.72  &0.65  &2.05  &0.1  &\ldots\      \\
18 &RGZ\_J082451.6+232254 &ARG000255z &126.21516 &23.38173 &39 &39 &1.0 &1 &1 &11.3 &49.16  &0.8   &1.97  &0.12 &\ldots\       \\
19 &RGZ\_J111147.3+232253 &ARG0002561 &167.94751 &23.38164 &38 &39 &0.97&1 &1 &18.9 &130.45 &0.73  &23.75 &0.17 &\ldots\   \\
\hline
\end{tabular}
\newline
The full, machine-readable version of this table (filename: `DR1\_FIRST\_radio\_classifications.csv') is available at the journal website and at Zenodo. A portion is shown here for guidance on form and content. 
\end{table}
}}

\clearpage
{\bf{
\begin{tiny}
  \begin{subtables}
    \begin{table}
      \caption{FIRST-based AllWISE infrared host properties (columns 1 to 11) of the first 10 lines of the host galaxy catalogue.}\label{dr1hosttab_FIRST_1}
      \begin{tabular}{ccccccccccc}
        \hline
        Cat ID & RGZ ID & Zooniverse ID & Host RA & Host Dec & AllWISE ID &WISE RA & WISE Dec & $W1$ & $W1$~Error & $W1$ SNR \\
        &        &               & deg     &deg       &            & deg    & deg      & mag  & mag        & \\
        (1) & (2) & (3) & (4) & (5) & (6) & (7) & (8) & (9) & (10) & (11) \\
        \hline
        5  &RGZ\_J143121.3+232251  &ARG0002565  &217.83901  &23.38094  &WISEAJ143121.35+232251.3 &217.83898 &23.38094 &14.49 &0.03 &39.0 \\
        6  &RGZ\_J140600.1+232249  &ARG0002566  & \ldots\   & \ldots\  &    \ldots\              &  \ldots\ & \ldots\ &\ldots\ &\ldots\ &\ldots\    \\
        7  &RGZ\_J105512.0+232306  &ARG000255x  &163.80027  &23.38534  &WISEAJ105512.05+232306.2 &163.80024 &23.38506 &15.73 &0.04 &25.0 \\
        8  &RGZ\_J084831.6+232245  &ARG0002567  & \ldots\   & \ldots\  &    \ldots\              &  \ldots\ & \ldots\ &\ldots\ &\ldots\ & \ldots\    \\
        9  &RGZ\_J171515.3+232247  &ARG0002568  &258.81392  &23.37965  &WISEAJ171515.35+232247.6 &258.81396 &23.3799  &16.66  &0.07 &14.7 \\
        10 &RGZ\_J164643.0+232254  &ARG000255v  &251.67925  &23.38219  &WISEAJ164643.03+232254.9 &251.67932 &23.38193 &15.45 &0.04 &27.8 \\
        11 &RGZ\_J091550.4+232254  &ARG000255y  &138.96021  &23.38203  &WISEAJ091550.46+232254.6 &138.96028 &23.38185 &15.97  &0.06 &18.7 \\
        15 &RGZ\_J095705.6+232253  &ARG0002560  &149.27349  &23.38161  &WISEAJ095705.65+232253.7 &149.27357 &23.38159 &15.69 &0.05 &22.8 \\
        18 &RGZ\_J082451.6+232254  &ARG000255z  &   \ldots\ & \ldots\  &    \ldots\              &  \ldots\ & \ldots\ &\ldots\ &\ldots\ & \ldots\    \\
        19 &RGZ\_J111147.3+232253  &ARG0002561  &167.94728  &23.38154  &WISEAJ111147.38+232253.7 &167.94744 &23.38159 &15.99 &0.06 &18.7 \\
        \hline
      \end{tabular}
    \end{table}

    \begin{table}
      \caption{Description of columns 12 to 23 for the AllWISE host galaxy catalogue.}\label{dr1hosttab_FIRST_2}
      \begin{tabular}{cccccccccccc}        
        \hline
        $W2$   & $W2$~Error & $W2$ SNR & $W3$   & $W3$~Error & $W3$ SNR &$W4$    & $W4$~Error & $W4$ SNR & $N_{\rm{MATCH}}^{\rm{WISE}}$ & $DWM$  & $z_{\rm{phot0}}$ \\
        mag    & mag        &          & mag    & mag        &          &mag     & mag        &          &                         &        &   \\
        (12)    & (13)       & (14)     & (15)   & (16)       & (17)     & (18)   & (19)       & (20)     & (21)                    & (22)   & (23) \\
        \hline
        14.20  &0.04       &29.7      &12.94  &0.48       &2.3       &9.38   &\ldots\     &$-$0.9      &1                        &\ldots\ &0.35785 \\
        \ldots\ &\ldots\     &\ldots\   &\ldots\ &\ldots\     &\ldots\   &\ldots\ &\ldots\     &\ldots\   &0                        &\ldots\ &\ldots\ \\
        15.76  &0.18       &6.0       &11.97  &\ldots\     &0.7       &8.52   &\ldots\     &0.4       &1                        &\ldots\ &0.41442 \\
        \ldots\ &\ldots\     &\ldots\   &\ldots\ &\ldots\     &\ldots\   &\ldots\ &\ldots\     &\ldots\   &0                        &\ldots\ &\ldots\ \\
        16.34  &0.19        &5.7      &12.66  &\ldots\     &0.2       &9.31   &\ldots\     &$-$1.2      &1                        &\ldots\ &\ldots\ \\
        15.27  &0.08       &14.4      &12.20  &\ldots\     &1.1       &9.16   &\ldots\     &$-$0.5      &1                        &\ldots\ &0.24539 \\
        15.58  &0.13       &8.6       &11.90  &\ldots\     &1.2       &8.79   &\ldots\     &0.1       &1                        &\ldots\ &0.98494 \\
        15.49   &0.11       &9.7      &12.11  &0.37       &2.9       &8.94   &\ldots\     &$-$1.2      &1                        &\ldots\ &0.50267 \\
        \ldots\ &\ldots\     &\ldots\   &\ldots\ &\ldots\     &\ldots\   &\ldots\ &\ldots\     &\ldots\   &0                        &\ldots\ &\ldots\ \\
        15.89  &0.16       &6.7       &12.59  &\ldots\     &$-$0.8    &8.84   &\ldots\     &0.1       &1                        &\ldots\ &1.00717 \\
        \hline
      \end{tabular}
      \newline
      The entire combined, machine-readable version of Tables 2a and 2b (filename: `DR1\_FIRST\_host\_properties.csv') is available at the journal website and at  Zenodo (DOI: 10.5281/zenodo.14195049). A portion is shown here for guidance on form and content.
    \end{table}
\end{subtables}

\end{tiny}

}}

\clearpage
{\bf{
\begin{table}
\caption{The first 10 lines of the ATLAS-based radio source morphology classification catalogue.}
\label{dr1morphtab_atlas}
\begin{tabular}{lllccccccccccccc}
\hline
Cat ID & RGZ ID & Zooniverse ID & RA & Declination & $N_{\rm{votes}}$ &$N_{\rm{total}}$ & CL & $N_{\rm{comp}}$ & $N_{\rm{peaks}}$ & LAE & TSA & OL & TF & TF err & DC \\
            &         &              &  deg & deg &             &              &    &               &               & arcsec & arcsec &mJy~beam$^{-1}$& mJy & mJy & \\
(1) & (2) & (3) & (4) & (5) & (6) & (7) & (8) & (9) & (10) & (11) & (12) & (13) & (14) & (15) & (16) \\
\hline
       5& RGZ J033133.0$-$283131  &  ARG0003r1f  &   52.88666 &   $-$28.52605   &      6   &      6 &  1.00   &    1 &      2&   98.7& 2025.73&5796.95 &2248.08 &   0.62 &      \ldots\ \\
       6& RGZ J032914.9$-$272332  &  ARG0003r1e  &   52.31236 &   $-$27.39255   &     35   &     38 &  0.92   &    1 &      2&   87.0& 2648.53&4986.10 &3085.17 &   0.61 &      \ldots\ \\
      12& RGZ J033323.7$-$272407  &  ARG0003r1q  &   53.34921 &   $-$27.40189   &      7   &      7 &  1.00   &    1 &      1&   57.4&  905.40&3796.80 &1101.89 &   0.27 &      \ldots\ \\
      13& RGZ J033409.2$-$282419  &  ARG0003r18  &   53.53858 &   $-$28.40529   &     25   &     35 &  0.71   &    1 &      1&   64.1& 1130.76&4170.21 &2288.64 &   0.33 &      \ldots\ \\
      14& RGZ J033405.2$-$282333  &  ARG0003r18  &   53.52216 &   $-$28.39259   &     25   &     35 &  0.71   &    1 &      1&   28.9&  255.60&4170.21 &  26.07 &   0.16 &      \ldots\ \\
      17& RGZ J033126.9$-$281811  &  ARG0003r1a  &   52.86145 &   $-$28.30475   &      6   &      6 &  1.00   &    1 &      1&  100.3& 3517.21&6296.00 &3094.92 &   0.88 &      \ldots\ \\
      18& RGZ J032846.5$-$282618  &  ARG0003r1b  &   52.19421 &   $-$28.43811   &     10   &     10 &  1.00   &    1 &      2&   64.2& 1102.68&3962.58 &2270.86 &   0.31 &      \ldots\ \\
      19& RGZ J033553.3$-$272740  &  ARG0003r1c  &   53.97216 &   $-$27.46153   &     24   &     32 &  0.75   &    1 &      1&   58.4&  991.80&3578.64 &2450.43 &   0.27 &      \ldots\ \\
      20& RGZ J033551.9$-$272652  &  ARG0003r1c  &   53.96716 &   $-$27.44745   &     24   &     32 &  0.75   &    1 &      1&   21.0&  134.64&3578.64 &   6.71 &   0.10 &      \ldots\ \\
      22& RGZ J033242.8$-$273817  &  ARG0003r1h  &   53.17691 &   $-$27.63808   &      7   &      7 &  1.00   &    1 &      2&   68.7& 1560.97&4436.65 &1660.41 &   0.41 &      \ldots\ \\

\hline
\end{tabular}
\newline
The full, machine-readable version of this table (filename: `DR1\_ATLAS\_radio\_classifications.csv') is available at the journal website and at Zenodo. A portion is shown here for guidance on form and content.
\end{table}
}}

\clearpage
{\bf{
\begin{tiny}
  \begin{subtables}
    \begin{table}
      \caption{ATLAS-based SWIRE infrared host properties (Columns 1 to 11) for the first 10 lines of the catalogue.}\label{dr1hosttab_ATLAS_1}
      \begin{tabular}{lllcclccccc}
        \hline
        Cat ID & RGZ ID               & Zooniverse ID & Host RA  & Host Dec  & SWIRE ID                   &SWIRE RA  & SWIRE Dec& $f_{\rm{3.6}}$ &$f_{\rm{3.6}}$ uncertainty & $f_{\rm{4.5}}$ \\
        &                      &               &deg       & deg       &                            &deg       &deg       &  $\mu$Jy     &$\mu$Jy                 & $\mu$Jy \\
        (1)    & (2)                  & (3)           & (4)      & (5)       & (6)                        & (7)      & (8)      & (9)          & (10)                    & (11) \\
        \hline
        5      &RGZ\_J033133.0$-$283131 &ARG0003r1f     &52.88768  &$-$28.52591  &SWIRE3\_J033133.02$-$283131.3 &52.88761  &$-$28.52537 &26.96         &0.86                     &30.46 \\  
        6      &RGZ\_J032914.9$-$272332 &ARG0003r1e     &52.31246  &$-$27.39279  &SWIRE3\_J032914.99$-$272332.6 &52.31250  &$-$27.39241 &54.71         &0.97                     &71.63 \\  
        12     &RGZ\_J033323.7$-$272407 &ARG0003r1q     &53.34941  &$-$27.40224  &SWIRE3\_J033323.78$-$272407.0 &53.34911  &$-$27.40197 &47.68         &1.05                     &59    \\  
        13     &RGZ\_J033409.2$-$282419 &ARG0003r18     &53.53887  &$-$28.40571  &SWIRE3\_J033409.29$-$282419.2 &53.53875  &$-$28.40535 &68.55         &1.1                      &42.74 \\  
        14     &RGZ\_J033405.2$-$282333 &ARG0003r18     &53.52206  &$-$28.39284  &SWIRE3\_J033405.25$-$282333.8 &53.52188  &$-$28.39274 &30.29         &0.91                     &36.59 \\  
        17     &RGZ\_J033126.9$-$281811 &ARG0003r1a     &52.86292  &$-$28.30352  &SWIRE3\_J033126.99$-$281811.2 &52.86250  &$-$28.30313 &481.66        &1.59                     &508.52\\  
        18     &RGZ\_J032846.5$-$282618 &ARG0003r1b     &52.19414  &$-$28.43877  &SWIRE3\_J032846.56$-$282618.1 &52.19404  &$-$28.43838 &70.11         &0.89                     &84.53 \\  
        19     &RGZ\_J033553.3$-$272740a&ARG0003r1c     &53.97227  &$-$27.46174  &SWIRE3\_J033553.33$-$272740.4 &53.97225  &$-$27.46123 &404.36        &2.05                     &366.98\\  
        20     &RGZ\_J033551.9$-$272652b&ARG0003r1c     &53.96689  &$-$27.44798  &SWIRE3\_J033551.90$-$272652.5 &53.96629  &$-$27.44793 &34.79         &0.64                     &26.72 \\  
        22     &RGZ\_J033242.8$-$273817 &ARG0003r1h     &53.17787  &$-$27.63825  &SWIRE3\_J033242.82$-$273817.6 &53.17845  &$-$27.63824 &45.49         &1.03                     &46.74 \\  
        \hline
      \end{tabular}
    \end{table}
    \begin{table}
      \caption{ATLAS-based SWIRE infrared host properties (Columns 12 to 19) for the first 10 lines of the catalogue.}\label{dr1hosttab_ATLAS_2}
      \begin{tabular}{cccccccc}
        \hline
        $f_{\rm{4.5}}$ uncertainty &$f_{\rm{5.8}}$ & $f_{\rm{5.8}}$ uncertainty & $f_{\rm{8.0}}$ & $f_{\rm{8.0}}$ &$N_{\rm{MATCH}}^{\rm{SWIRE}}$ & $z_{\rm{sp}}$ & $z_{\rm{ph}}$ \\
        $\mu$Jy                 & $\mu$Jy     & $\mu$Jy                 & $\mu$Jy      & $\mu$Jy      &                         &            &              \\
        (12)                     & (13)        & (14)                    & (15)         & (16)         & (17)                    & (18)       &(19)          \\
        \hline
        1.11                      & \ldots\     &\ldots\                  &\ldots\       &\ldots\       &1                        &\ldots\     &  \ldots\     \\                          
        1.02                      &109.28       &3.89                     &158.46        &3.77          &1                        &\ldots\     & \ldots\      \\       
        1.32                      &47.38        &4.86                     &\ldots\       &\ldots\       &1                        &\ldots\     & \ldots\      \\                  
        0.64                      &\ldots\      &\ldots\                  &\ldots\       &\ldots\       &1                        &\ldots\     &0.69          \\                
        1.16                      &\ldots\      &\ldots\                  &\ldots\       &\ldots\       &1                        &\ldots\     &  \ldots\     \\                          
        2.45                      &567.93       &4.41                     &750.78        &5.18          &1                        &\ldots\     &0.236         \\
        0.81                      &102.19       &3.84                     &83.92         &2.88          &1                        &\ldots\     &  \ldots\     \\       
        2.16                      &286.58       &5.32                     &959.02        &5.69          &1                        &\ldots\     &   \ldots\    \\     
        0.77                      &\ldots\      &\ldots\                  &\ldots\       &\ldots\       &1                        &\ldots\     &  \ldots\     \\                          
        1.24                      &\ldots\      &\ldots\                  &65.42         &5.06          &1                        &\ldots\     &1.148         \\            
        \hline 
      \end{tabular}
      \newline
      The entire combined, machine-readable version of Table 4a and 4b (filename: `DR1\_ATLAS\_host\_properties.csv') is available at the journal website and at  Zenodo (DOI: 10.5281/zenodo.14195049). A portion is shown here for guidance on form and content.
    \end{table}
  \end{subtables}  
\end{tiny}
}}

\end{landscape} 

\clearpage

\subsection{Radio properties of DR1 sources}
The main strength and core outcome of the RGZ project is the classification of radio source morphologies 
through the mutual association of discrete radio components.
Of the 99,146 FIRST radio sources (from 99,602 catalogue entries) presented, 16,354 DR1 radio sources are composed of more than one  
component. DR1's FIRST-based catalogue of visually-classified radio morphologies is greater than than that from 
the Combined NRAO VLA Sky Survey \citep[NVSS; ][]{condon98}-FIRST Galaxies \citep[CoNFIG; ][]{gendre08} sample by two orders of magnitude. The number of 
multicomponent radio sources presented by DR1 is of the same order of magnitude as radio source samples
that have been compiled via automated algorithms such as those from \citet{proctor11,vanvelzen15}, in addition to recent surveys such as \citet{williams19}.
The advantage of the DR1 catalogue over that of \citet{vanvelzen15} is that we are able to classify radio sources which have angular
extents larger than 1\arcmin, and that all classifications have been visually-inspected. Such visual inspection is important; e.g.,  visual host-galaxy identifications for the G4Jy Sample \citep{white20b,white20a}, found discrepancies for several very bright sources that overlapped with the catalogue of \citet{vanvelzen12}.

\subsubsection{Angular sizes of multicomponent sources}
\begin{figure}
\begin{center}
\includegraphics[scale=.5]{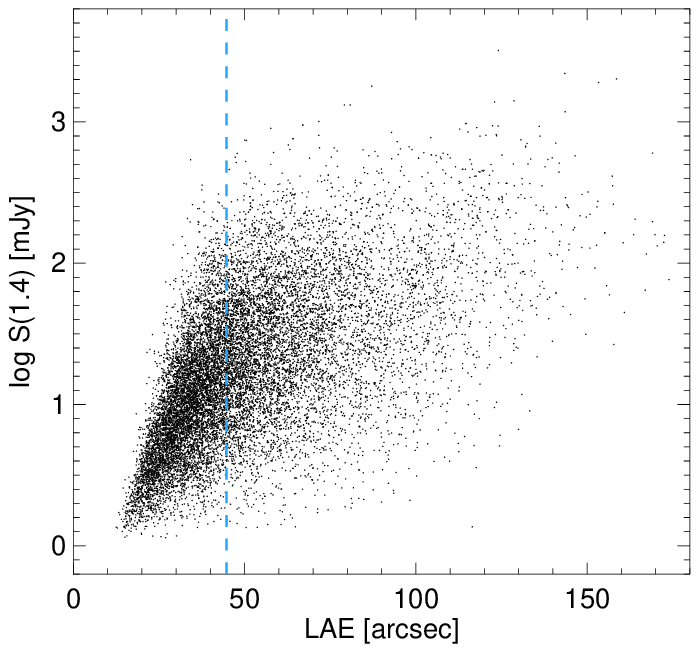}
\caption{Integrated flux as a function of largest projected source angular extent for RGZ DR1 sources with more than one radio component.  The dashed blue line marks the median angular size.}
\label{flxsize}
\end{center}
\end{figure}
The angular size of each source is measured here by the largest angular extent ($LAE$; Section~3.1) parameter
 in the catalogue.   The RGZ DR1 FIRST sample consists of 5,310 radio sources larger than 1~arcmin.  Figure~\ref{flxsize} shows the total 1.4 GHz flux ($TF$) as a function of $LAE$ for the RGZ DR1 sources with more than one radio component.  In a similar study, \citet{magliocchetti98}  found two distinct populations when comparing the sizes and total fluxes of double sources (which could be sources with two components or two peaks in DR1) from the FIRST survey.  In \citet{magliocchetti98}, the second concentration of double sources with extents greater than 1~arcmin are thought to consist of unrelated pairs of sources.  Since Figure~\ref{flxsize} does not show a secondary population of extended sources with angular sizes greater than 1~arcmin, we conclude that DR1's extended radio source sample is relatively free of such misclassifications.  Our result is consistent with recent findings from the VLASS survey which suggest that double sources with sizes that are larger than 100~arcsec are likely to be unrelated pairs of sources \citep{gordon23}.

\begin{figure}
\begin{center}
\includegraphics[scale=.5]{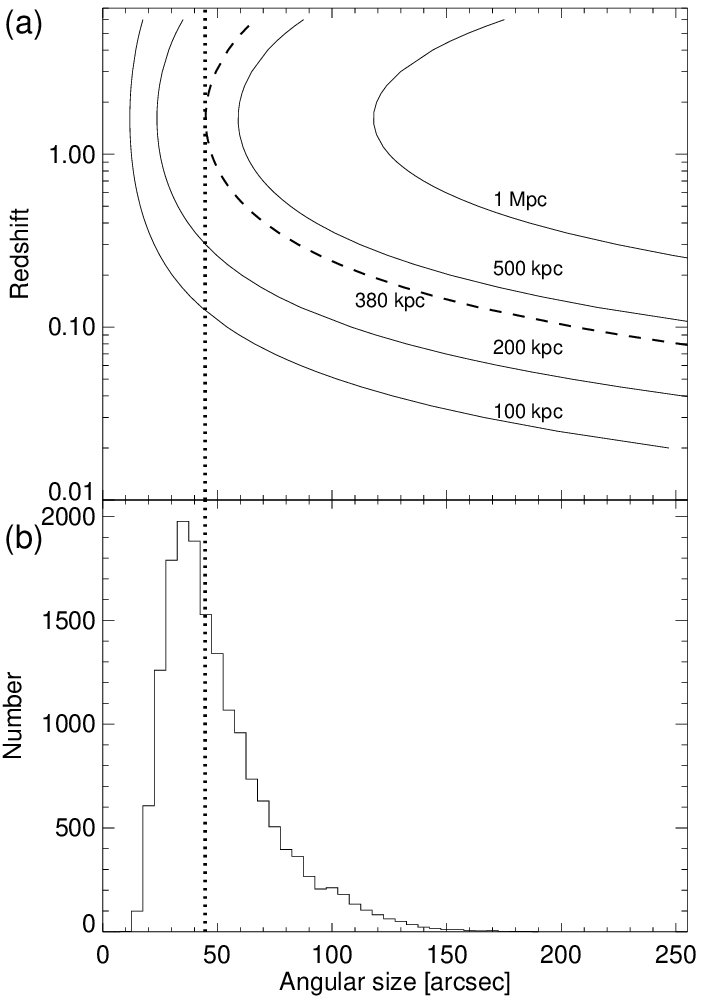}
\caption{Panel (a): The observed angular size as a function of redshift assuming the cosmology 
constants described in Section 1. Constant lines of angular size are plotted
with solid lines and labelled accordingly. The dashed line represents the constant linear size of 380~kpc, which is the largest linear size, at any redshift, that is consistent  with the median angular size of the multicomponent DR1 sources. This result suggests that at least half of the multicomponent DR1 sources have physical sizes that are smaller than 380~kpc.    Panel (b): Distribution of angular extents for DR1 multicomponent sources. The dotted vertical line in both panels mark the median LAE at 44.7\arcsec.}
\label{maedist}
\end{center}
\end{figure}

\begin{figure}
\begin{center}
\includegraphics[scale=.4]{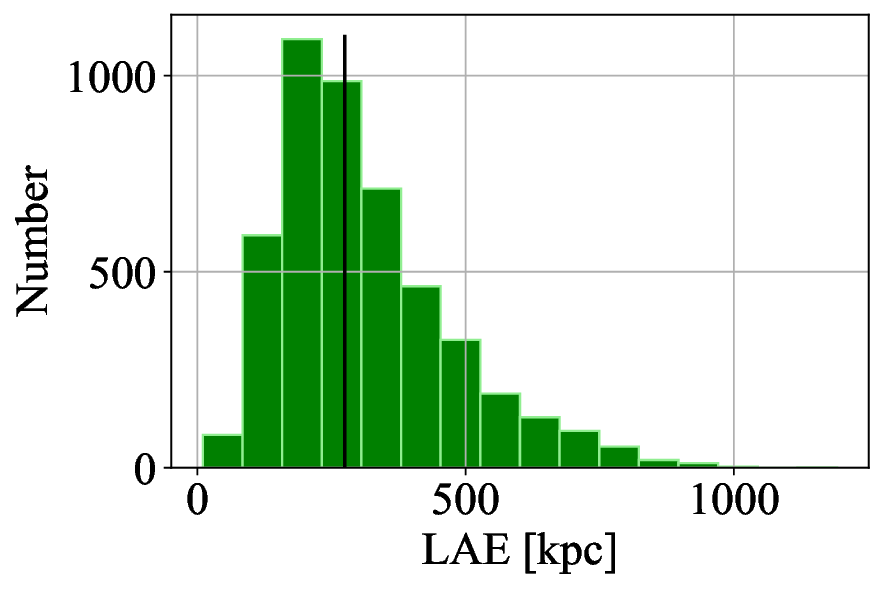}
\caption{Distribution of physical LAE for multicomponent DR1 FIRST sources based on photometric redshifts from \citet{beck22}.
The vertical black line marks the median LAE of 275~kpc.}
\label{laephys_dist}
\end{center}
\end{figure}


We show the distribution of $LAE$ for the 11,092  multicomponent radio sources
in Figure~\ref{maedist}b. The median $LAE$ is 44.7\arcsec (represented by the vertical dotted line in 
Figure~\ref{maedist}).  Only 56~percent of the FIRST-based DR1 classifications are matched to WISE host galaxies with $W1$~SNR greater than or equal to 4.0.  We deduce  from the median angular size in both panels (a) and (b) of Figure~\ref{maedist} that at least half of the multicomponent radio sources in DR1 are likely to have physical extents that are smaller than 380~kpc (represented by the dashed line in Figure~\ref{maedist}a).  Following the recent publications of large photometric redshift catalogues \citep[e.g.\ ][]{bilicki16,beck22,dalya22,duncan22}, we cross-matched our WISE host galaxies to that of \citet{beck22} to determine the projected physical extents of the DR1 multicomponent radio sources from the estimated redshifts.  Consistent with our estimation from Figure~\ref{maedist}, we find that the DR1 multicomponent radio sources have a mean and median projected LAE of 311~kpc and 275~kpc, respectively (Figure~\ref{laephys_dist}). 

 We note that at low redshifts, there may be an additional contribution of smaller sources with angular extents
 that are larger than the observed median angular size.  Some local Universe sources with large angular extents
 may also be missed or inaccurately classified due to the 3\arcmin\ size of the RGZ subjects.

 Conversely, any large radio source with a physical extent of 500~kpc corresponds to an observed
 minimum angular extent of approximately 1~arcmin (Figure~\ref{maedist}a) at redshifts between 1 and 2.
 Giant radio galaxies \citep[e.g.\ ][]{ishwara99} are typically defined to have minimum extents of
 700~kpc \citep{dabhade17,kuzmicz18} to 1~Mpc \citep{andernach12,andernach21}. 
 Hence, the 3\arcmin\ field-of-view of each RGZ subject is sufficient for the identification of such sources
 at redshifts between 0.2 and 5.0. A description of giant radio galaxies from RGZ DR1 can be found in \citet{tang20}.

\subsubsection{Extended radio sources with no infrared counterparts}
Infrared faint radio sources (IFRS) are thought to reside at cosmological redshifts \citep{norris06,collier14,herzog14,orenstein19,patil19,white20a}.
In DR1, there are 17,025 FIRST-based extended radio sources (i.e.\ sources with more than one radio peak) with no WISE counterpart, hereafter referred to as no-IR sources.  

\begin{figure}
\includegraphics[scale=.4]{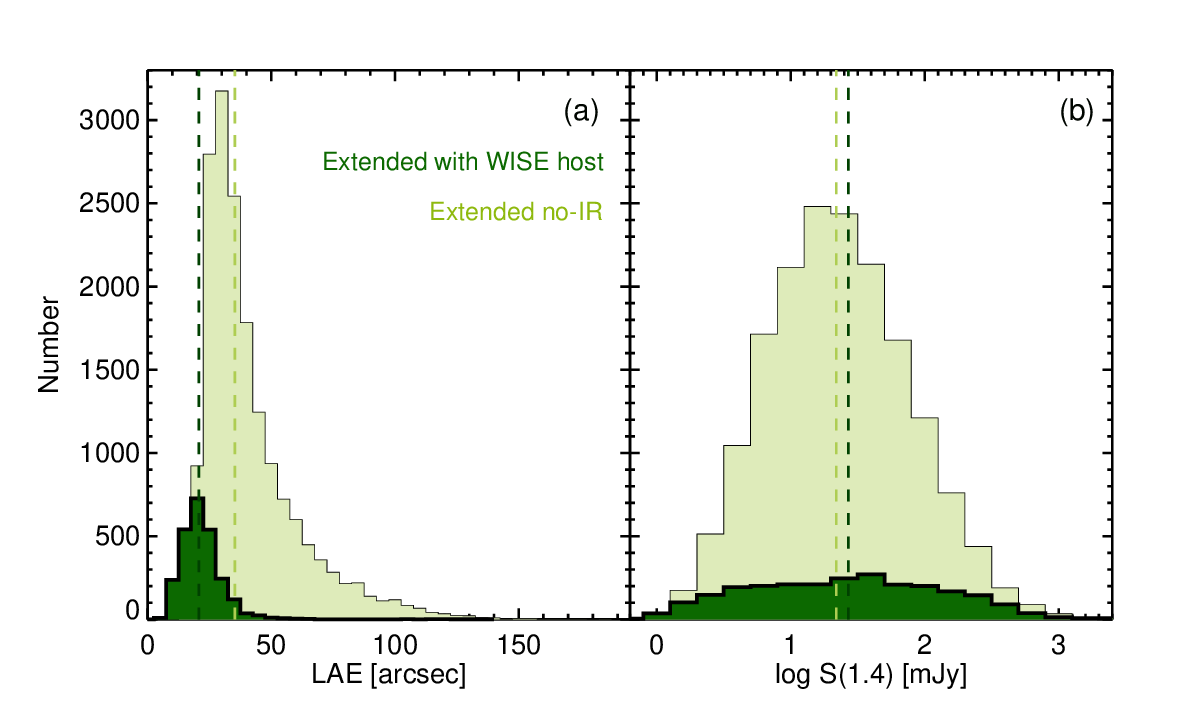} \\
\caption{Distributions of the LAE (panel a) and total flux density (panel b). No-IR extended (multiple-peak) radio sources  are shaded in light green, while  extended (multiple-peak) sources with matched WISE host galaxies are shaded in dark green. The no-IR extended radio sources are typically larger in angular size than those matched with WISE host galaxies.  The light and dark green vertical dashed lines mark the median values for the no-IR and radio sources with known hosts, respectively.  }
\label{laenoir_compare}
\end{figure}

How similar are extended no-IR sources to extended radio sources with matched WISE hosts?  While the two populations of extended radio sources show similar total flux distributions, we find that the $LAE$ for extended no-IR sources and extended sources with hosts are significantly different (Figure~\ref{laenoir_compare}). A two-sided Kolmogorov-Smirnov (KS) test of the two sample-normalised $LAE$ distributions finds a KS $p$-value of 0.005.  The median $LAE$ for the no-IR sources (35~arcsec) is also greater than that of the sources with matched WISE hosts (21~arcsec).    While larger objects are typically nearer, we know that the sensitivity of WISE limits us to a view of the MIR Universe to redshifts below one \citep{yan13,jarrett17} and so if the larger radio sources are nearer, the host galaxy should be apparent in the WISE observations. The larger median $LAE$ of the no-IR sources therefore suggests that these sources could be more powerful radio galaxies at higher redshifts \citep[e.g.\ ][]{bruggen21,delhaize21}.

\begin{figure}
\includegraphics[scale=.4]{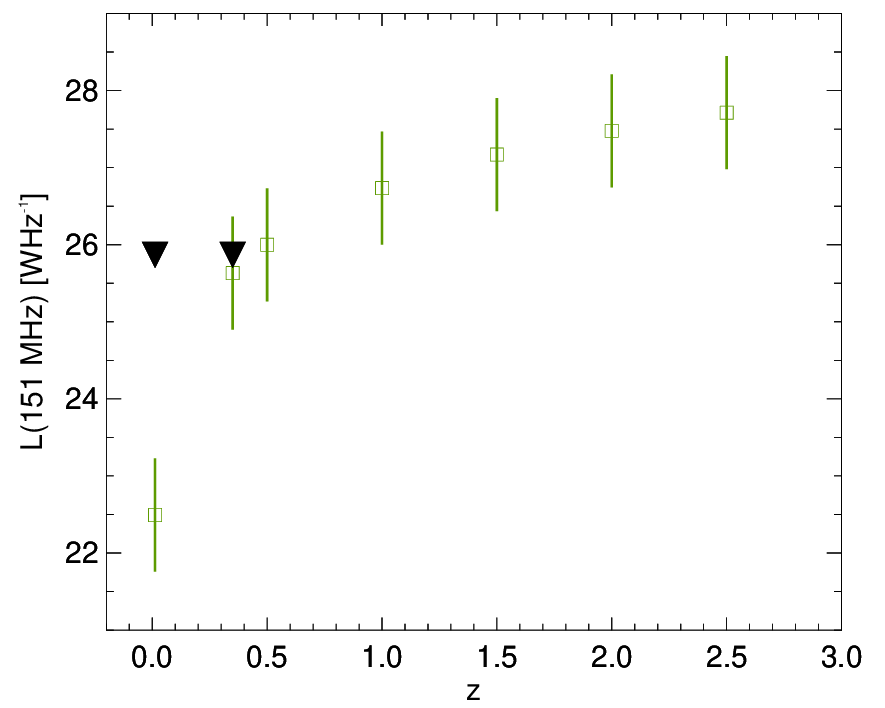} \\
\caption{The upper limit estimates of 151~MHz luminosities from scaling the $W1$ upper limit to distances of $z=0.012$ ($\approx$50~Mpc) and $z=0.35$ (solid black inverted triangles). These upper limit estimates are then compared to the median (green squares), and interquartile ranges (vertical errorbar), of the 151~MHz luminosities for the sample of extended no-IR sources.  The green squares and interquartile ranges are estimates which assume the $W1$ detection limit and the assumption that the extended no-IR sample all reside within each of the 7 fiducial redshift bins. }
\label{lumnoir_compare}
\end{figure}

To constrain the redshift of the extended no-IR DR1 sample, we combine the WISE $W1$ flux limit of 54~$\mu$Jy \citep{cutri13} with known scaling relationships between galaxy stellar masses \citep{wen13}, black hole masses \citep{reines15} and the 151~MHz luminosities \citep{meier01,godfrey13}.  Consistent with the limits on the stellar mass scaling relationships described by \citet{wen13}, we estimate the upper limits of the 151-MHz luminosities at redshifts 0.012 (corresponding to a distance of 50~Mpc) and 0.35 from the $W1$ detection limit (represented by black inverted triangles in Figure~\ref{lumnoir_compare}).  For comparison, we place the sample of DR1 no-IR extended sources at 7 fiducial redshifts, and estimate the 151~MHz luminosities from the total 1.4~GHz fluxes ($TF$) by assuming a power law spectral index of $\alpha=-0.7$ (where $S_{\nu} \propto \nu^{\alpha}$).  The medians of these fiducial 151~MHz luminosities and interquartile ranges are represented by the green open squares and errorbars in Figure~\ref{lumnoir_compare}. In this comparison, we find that the no-IR sample is unlikely to be located in the local Universe (50~Mpc) because the upper limits inferred from a $W1$ non-detection are much greater than that estimated by the median fiducial 151~MHz if we placed the entire no-IR sample at a distance of 50~Mpc. The inferred low luminosities (at 50~Mpc) are not consistent with the typical luminosities of extended radio galaxies.  The median upper interquartile range of estimated 151~MHz luminosities for no-IR sources at $z=0.35$ is consistent with the luminosity upper limit derived from the $W1$ detection limit.  Similarly,  the median of the estimated 151~MHz luminosities assuming $z \geq 0.5$ is consistent with luminosities that are expected for extended radio-loud galaxies \citep[e.g.\ ][]{sejake22}.  As such, the FIRST DR1 no-IR sample of 17,025 sources may be a useful starting point for future follow-up studies of extended radio galaxies at intermediate redshifts ($z \geq 0.35$).

\begin{figure}
\includegraphics[scale=.4]{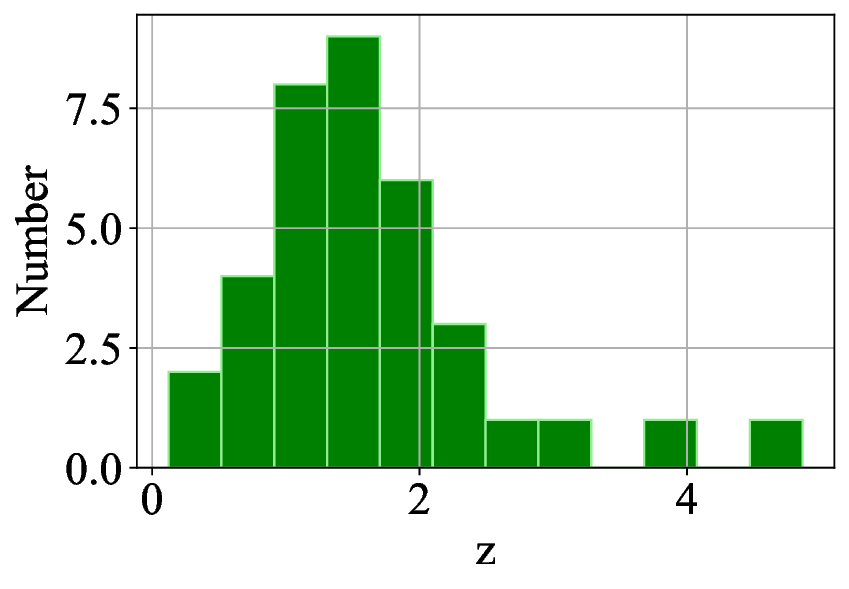} \\
\caption{Redshift distribution of the cross-matched extended no-IR sample to LoTSS-deep DR1 \citep{duncan21}. }
\label{lotss_compare}
\end{figure}

Similar to Section~5.1.1, we cross-matched the radio source positions of the extended no-IR sample with that of the LOFAR Two-metre Sky Survey deep fields Data Release 1 \citep[LoTSS-deep DR1; ][]{duncan21} and found a mean redshift of 1.584 for our cross-matched sample of 38 extended no-IR sources.  Figure~\ref{lotss_compare} shows the redshift distribution from this cross matching with LoTSS-deep DR1, suggests that there are 2 cross-matched RGZ DR1 sources that reside at redshifts lower than 0.35.  While one of these (RGZ\_J142557.8+351258) has a `probable' quality flag to the catalogued spectroscopic redshift estimate of 0.123, the other source (RGZ\_J143100.4+353631) has a reliable spectroscopic redshift measurement of 0.162. We note that the fraction of lower redshift interlopers (two of 38) within the no-IR sample is consistent with the expected reliability of DR1.

\subsection{DR1 host galaxy properties}

\begin{figure}
\begin{center}
\begin{tabular}{c}
\includegraphics[scale=.4]{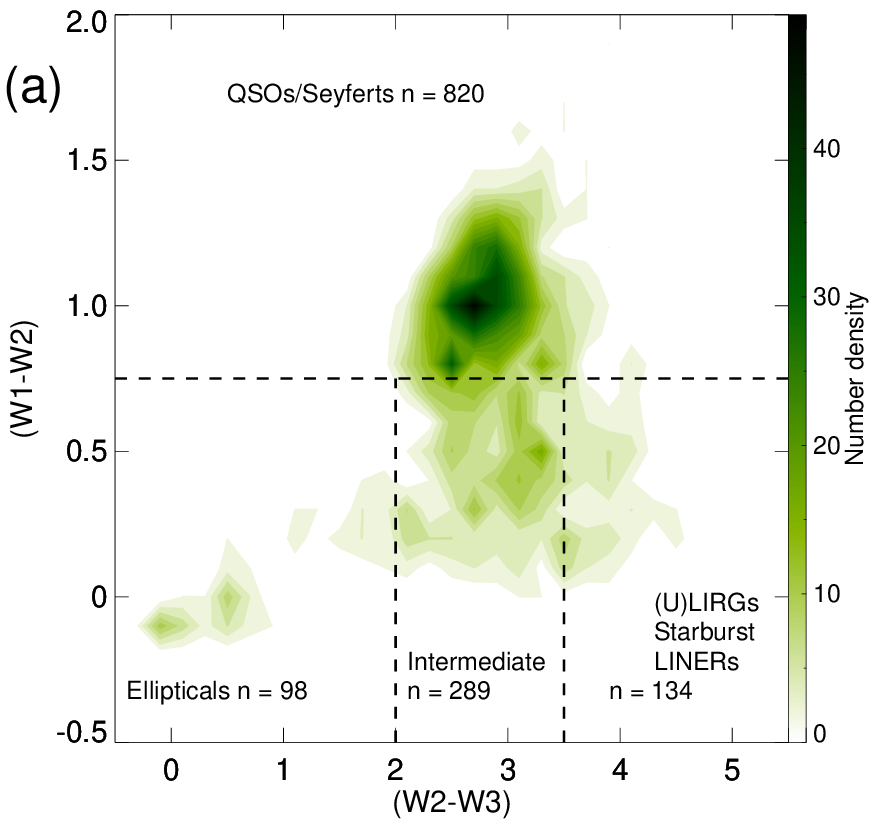} \\
\includegraphics[scale=.4]{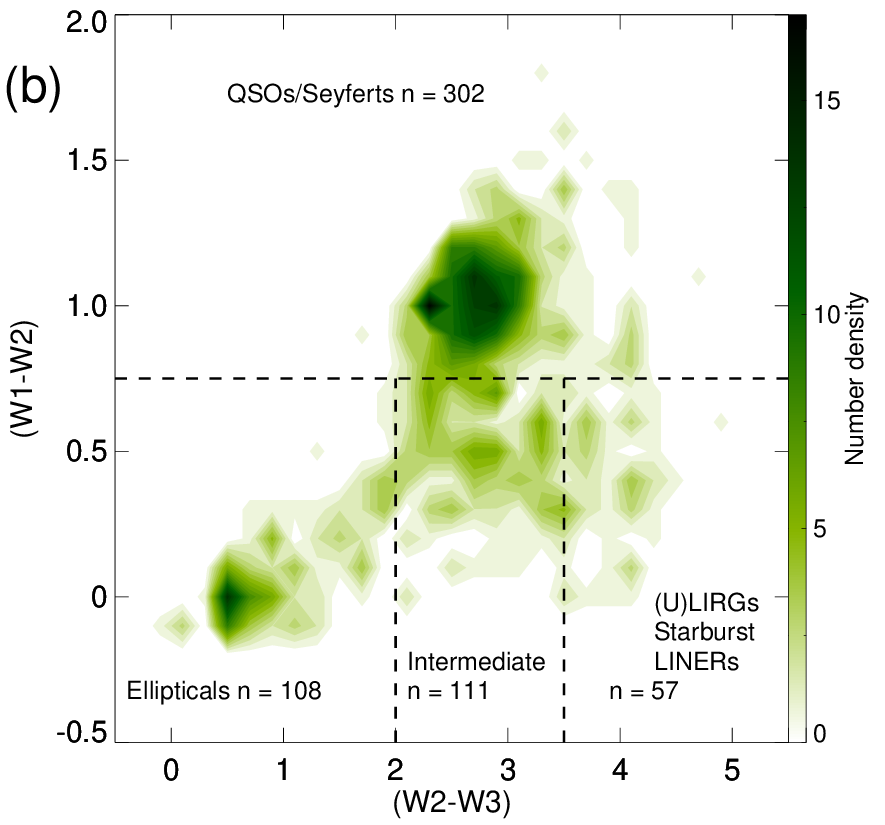} \\
\end{tabular}
\end{center}
\caption{WISE colour-colour diagram for RGZ sources with small ($<44.7$\arcsec; panel a) and the largest top 20~percent of angular extents of the 2,888 extended sources ($N_{\rm{peaks}} >1$; panel b). The colour scale represents the number density per area where the area is defined to have colour bin sizes of $(W1-W2)=0.1$ and $(W2-W3)=0.2$.}
\label{wisecolcol_mae}
\end{figure}

The RGZ DR1 catalogue includes the WISE $W1$, $W2$, $W3$ and $W4$ magnitudes of the cross-matched WISE host galaxies.  A total of 7,656 FIRST-based DR1 sources  are matched to host galaxies with WISE $W1$, $W2$ and $W3$ detections with a minimum of 4-sigma significance in each band. Consistent with previous results \citep[e.g.\ ][]{kurcz16}, the signal-to-noise requirement for $W3$ has eliminated $\approx86$~percent of the sources from the parent DR1 FIRST catalogue that have been matched to AllWISE sources detected in $W1$ (at $>4\sigma$).  The $W1-W2$ versus $W2-W3$ distribution of DR1 sources with matched WISE hosts indicates that the DR1 sample is consistent with the MIR colour distributions typical of the QSO-Seyfert population, the elliptical galaxies, the star-forming (LIRGs, starburst or LINERS) population, in addition to the intermediate colour region between that of the elliptical and star-forming populations \citep{wright10,jarrett17,hardcastle19}. \citet{jarrett17} also refers to the intermediate colour region as the ``intermediate disk" region within the MIR colour distribution. The dashed lines in Figure~\ref{wisecolcol_mae} mark the location of these MIR colour-colour regions that are dominated by the different population of sources at low redshifts \citep{jarrett17,alger21}.

Of the 7,963 DR1 sources (with SNR$\geq4$ in $W1$, $W2$ and $W3$ simultaneously), 2,888 are extended and have $N_{\rm{peaks}} > 1$. We divide this extended sample of multipeak radio sources into two classes: one with $LAE$ less than the median $LAE$ of 44.7~\arcsec (1,341 sources); and another with  $LAE$ that is the largest 20~percent of the sample (578 sources with a minimum $LAE$ of 72.9~arcsec).   Figure~\ref{wisecolcol_mae} shows the MIR colour distribution for extended radio sources with small (panel a) and the largest angular extents (panel b). Typically, the MIR colours for these extended sources across all angular sizes are dominated by colours that typify QSOs. The relative fraction of multicomponent sources with WISE elliptical colors (relative to QSO colors) is a factor of three larger for the larger sources, compared to the smaller ones.  We note that the MIR colour regions delineated by the dashed lines in Figure~\ref{wisecolcol_mae} are relevant to galaxies residing at low redshifts.  As we do not have spectroscopic redshifts for our sample, the sources within the MIR colour-colour regions will likely be contaminated with higher redshift sources which have been redshifted into a different colour region away from their originating MIR colours \citep[e.g.\ ][]{donley12,gurkan14,mingo16,alger21,gordon23}.  Furthermore, the initial RGZ FIRST sample selection for extended radio sources likely affects the distribution of MIR colours that are observed.  For example many bright and compact radio sources will not be included in the RGZ FIRST sample.

\subsection{Comparing the radio and MIR fluxes}
Previous studies by \citet{mingo16} have found  the 1.4~GHz flux density -- $W3$ 
magnitude parameter space to be an effective discriminator between radio
continuum emission arising from star formation versus a radio AGN.
Here we compare the total 1.4~GHz flux densities  to the $W3$ magnitude
of the host galaxy for single-component and extended sources in Section~5.3.1 and
Section~5.3.2, respectively. The analysis in this section is based 
on the subsample of 7,963 DR1 sources  where reliable MIR magnitudes are available (SNR$\geq4$ in $W1$, $W2$ and $W3$ simultaneously, as per Section~5.2). Of the 7,963 sources with reliable MIR observations, 2,888 sources are extended.

\subsubsection{Single component compact radio sources}
\begin{figure}
\includegraphics[scale=.3]{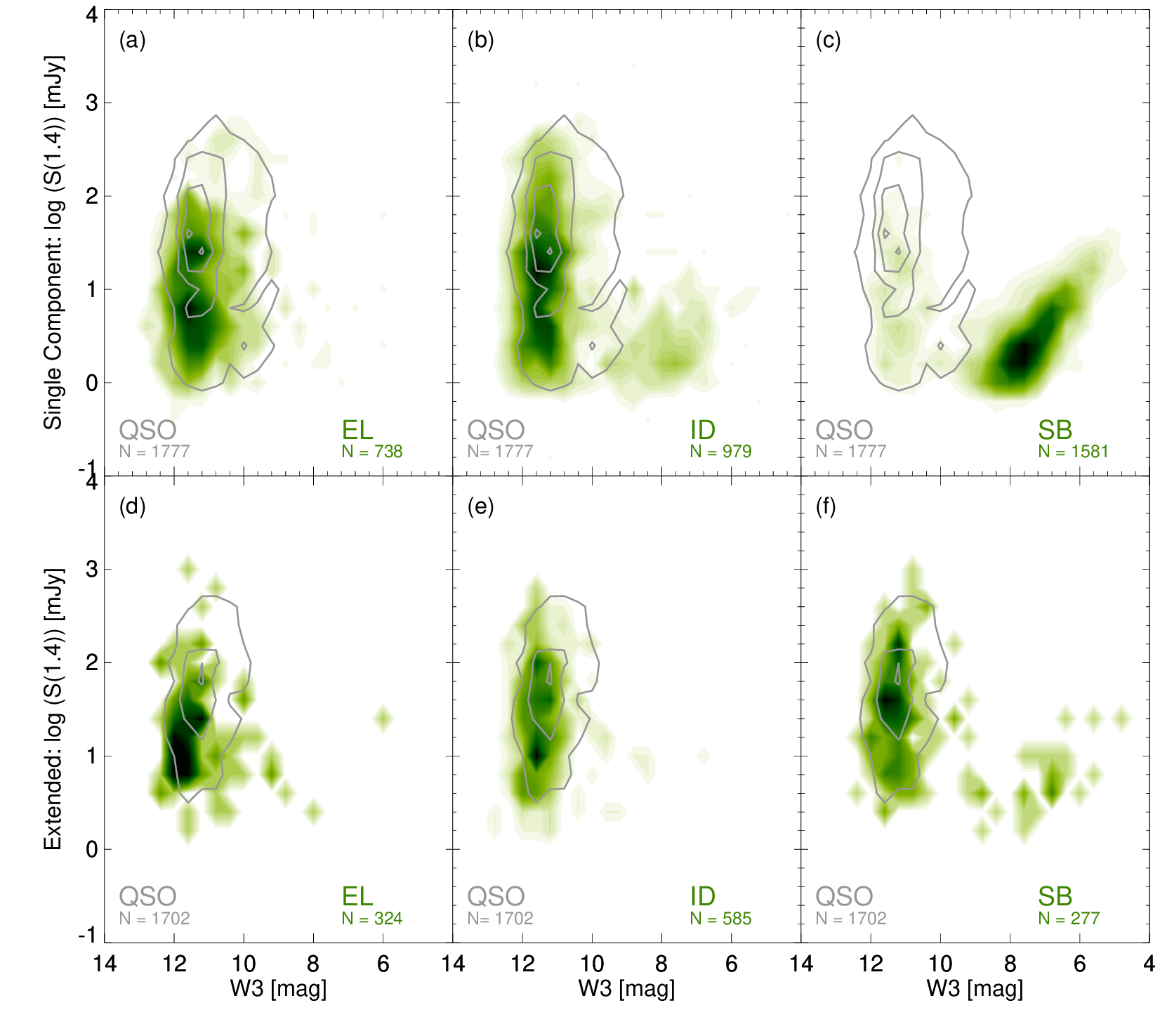} \\
\caption{Distributions of 1.4 GHz total flux densities as function of W3 (12~$\mu$m) magnitudes for single component sources (panels a, b and c) and extended sources (panels d, e and f). Panels (a) and (c) show the single component and extended sources with  elliptical (EL)  MIR colours in green, respectively. Panels (b) and (d) show the single component and extended sources with intermediate (ID) MIR colours in green, respectively. Panels (c) and (f) show the single component and extended sources with starburst-LINER (SB) MIR colours in green, respectively. The single-component QSO population ($N=1777$) is overlaid as grey contours in  panels (a), (b) and (c). The extended QSO population ($N=1702$) is overlaid as  grey contours in panels (d), (e) and (f). The green colour scale and grey contours represents the number density per area where the area is defined to have binsizes of $W3=0.4$~magnitude and $S_{1.4}=0.2$~mJy.}
\label{flxmag_sc}
\end{figure}

The top row (panels a, b and c) of Figure~\ref{flxmag_sc} presents the 1.4~GHz total flux densities 
as a function of the host $W3$ magnitude for 5,075 single component
sources that have been classed according to the WISE colour-colour 
classes marked by the dashed lines in Figure~\ref{wisecolcol_mae}. In the top row
 of Figure~\ref{flxmag_sc}, the 
distribution of single component sources with QSO MIR colour properties is overlaid
as grey contours in each panel.  The green shaded regions in panels (a),
(b), and (c) represent the single component sources with elliptical, intermediate
disks and star formation  MIR colour properties, respectively.  Consistent
with \citet{mingo16}, we find that the single component sources with star formation
MIR colours lie offset to brighter $W3$ magnitudes away from the $W3$ magnitudes that
typify the population of ellipticals and QSO, and also show the classic radio-IR correlation 
that is typical of star-forming galaxies.
On the other hand, there remains a small sub-population of starburst/LINER-associated 
sources which exhibit similar 1.4~GHz flux density -- $W3$ magnitudes
as those of ellipticals and QSOs --- suggesting that an AGN origin may also be possible
in addition to star formation.

The  DR1 single component sources with intermediate disk MIR colours have 
a larger fraction of 1.4~GHz flux densities and $W3$ magnitudes that 
are consistent with the distributions found for the QSO and elliptical
population and a small fraction which exhibit star-forming properties
(Figure~\ref{flxmag_sc}b). Despite having bluer $W1-W2$ colours, the majority of 1.4~GHz emission 
from intermediate MIR colour region may originate from  faint 
radio-quiet AGN rather than star formation.  This hypothesis is supported by
 studies that have demonstrated that radio-quiet AGN are fairly common
 in many galaxies \citep[e.g.\ ][]{white15,wong16,white17}.  

\subsubsection{Extended sources}

We show the 1.4~GHz flux densities and  $W3$ magnitudes
for 2,888  extended sources in the bottom row of Figure~\ref{flxmag_sc} (panels d, e and f).  Similar to 
Figure~\ref{flxmag_sc}a, b and c, the extended sources with
elliptical, intermediate disk and starburst-LINER MIR colors are represented by the green shaded regions in Figure~\ref{flxmag_sc}d, e and f, respectively.  The QSO distribution is also overlaid as grey contours in each panel.  
The distributions of 1.4~GHz flux densities and 
$W3$ magnitudes for extended radio sources are nearly identical for those of the QSO, intermediate disk and elliptical populations (Figure~\ref{flxmag_sc}d and e), suggesting a common AGN origin for these
extended radio sources found in the DR1 FIRST sample.  We also note that there is
a small number of extended radio sources which exhibit MIR colours that are typical of starbursts or LIRGs (Figure~\ref{flxmag_sc}f). 

\subsection{Limitations of this data release}

The main product from this  data release is the  visual classification
 of 99,146 radio sources from FIRST and 583 sources from ATLAS with classification 
consensus levels greater than or equal to 0.65.  This section has demonstrated the science-readiness
of RGZ DR1, one of the largest visually-classified catalogues of radio morphologies. 
Further demonstration of the scientific use of 
RGZ DR1 classifications come from recent successes in the development of more advanced deep learning-based methods for automated radio source classifications.

On the other hand, we note that the $LAE$ and the integrated fluxes may be underestimated for the DR1 FIRST sample \citep{wu19} as the FIRST survey is less sensitive to extended diffuse radio emission than NVSS.  The latter is a survey undertaken with the same telescope but with a shorter baseline array configuration which, trades-off angular resolution for surface brightness sensitivity. For multi-component sources with more compact hotspots (FR-II sources), the $LAE$ can also be overestimated by approximately 15~percent.  We refer the reader to \citet{white97} for a more detailed comparison of the radio photometry from FIRST relative to NVSS.

\begin{figure}
\includegraphics[scale=.7]{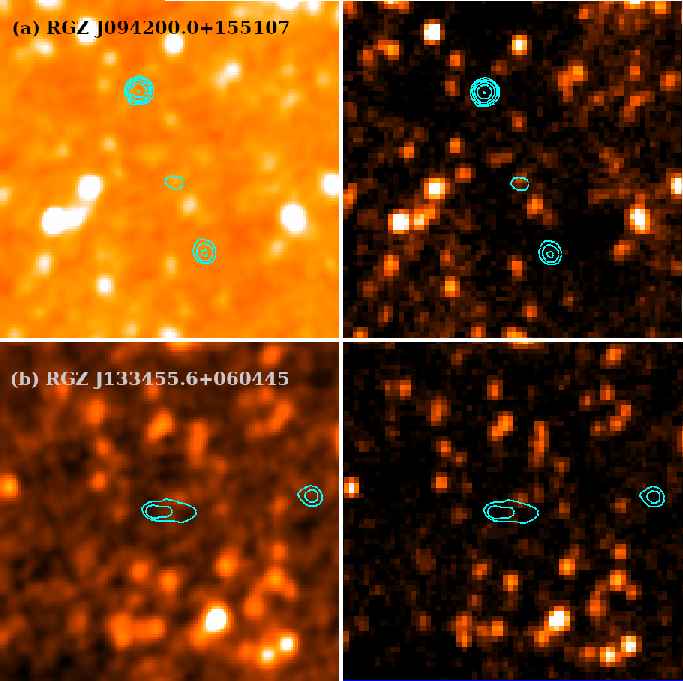}
\caption{Example comparisons of AllWISE (left column) and unWISE (right column) $W1$ observations for two RGZ sources: RGZ\_J094200.0+155107 (row a) and RGZ\_J13455.6+060449 (row b). The FIRST radio contours are overlaid in cyan and start at the 4$\sigma$ level and increase by factors of $\sqrt{3}$.}
\label{neowise}
\end{figure}

Since the completion of the project, deeper WISE and NEOWISE $W1$ images are now available to the community, such as the 8-year unWISE coadds \citep{meisner22,marocco21}.  As such, it is likely that the number of no-IR sources that are currently identified can be reduced with such enhancements in sensitivity of the IR images.  We provide two example RGZ subject comparisons of the AllWISE and the deeper unWISE $W1$ images in Figure~\ref{neowise}.  The example subject in Figure~\ref{neowise}a demonstrates the value of more sensitive $W1$ observations in revealing an IR host galaxy.  We contrast this finding with that of  Figure~\ref{neowise}b where the increase in source density of the deeper unWISE image did not reveal a more prominent host galaxy, despite the increase in source density.

Pre-release versions of DR1 have been used by several team members as input 
training sets for advanced deep learning algorithms 
in order to further automate the classification of radio source morphologies \citep[e.g.\ ][]{galvin19,ralph18,tang22,slijepcevic22}.  Such automated classification methods are necessary for the
very large number of radio sources that we expect from the SKA era of surveys \citep[e.g.\ ][]{norris21,gupta24}.  However, we acknowledge that visual verification and classification may be required for the rarer and more complex radio morphologies.  As such, automated classifications will help with reducing the number of sources that require manual inspection, thereby improving the discovery efficiency of future generations of Radio Galaxy Zoo-like citizen projects \citep[e.g.\ ][]{walmsley23}.  


\section{Summary and conclusions}
Radio Galaxy Zoo Data Release 1 presents one of the largest catalogue of quantified visual classifications of
radio source morphologies to-date. This data release consists of 100,185 radio classifications
for 99,146 radio sources from FIRST survey and 582 radio sources
from the ATLAS survey, with an average reliability of 0.83.  Cross-identifications of the host galaxy
with mid-IR objects from the AllWISE and SWIRE surveys are presented for 55,731 and 502 catalogue entries from the FIRST and ATLAS surveys, respectively.

The FIRST-based radio sources show MIR colour properties which are well distributed across the MIR colour regions populated by QSOs, intermediate disk galaxies, elliptical galaxies and starbursts or LINER sources. On the other hand, a comparison of the 1.4~GHz flux to the $W3$ magnitude finds good consistency between the sources that have MIR colour properties similar to ellipticals and QSOs. There are 17,025 extended no-IR sources within RGZ DR1 which have a significantly different distribution of LAE (median of 35~arcsec) relative to that of extended DR1 sources with matched WISE host galaxies (median of 21~arcsec). We argue that these no-IR sources reside at redshifts beyond 0.35, possibly related to the population of higher redshift IFRS.




\section*{Acknowledgements}

This publication has been made possible by the participation of more than 12,000 volunteers in the Radio Galaxy Zoo project.  Their contributions are individually acknowledged at http://rgzauthors.galaxyzoo.org.  We thank the anonymous referee and scientific editor for improving the clarity of this paper.  We also acknowledge S.~George, M.~Gendre, Y.~Gordon, A.~Kapadia, E.~Middelberg, R.~Simpson, A.~Smith, C.~Snyder, J.~Tate   and C.~Wu who have made valuable contributions to this paper and project. 
Partial support for this work for AFG, KWW and LR is provided by the U.S.\ National Science Foundation grants AST-112595 and AST17-14205 to the University of Minnesota.   SS thanks the Australian Research Council for an Early Career Fellowship DE130101399. HA has benefited from University of Guanajuato grant DAIP \#66/2018. HT gratefully acknowledges the support from the Shuimu Tsinghua Scholar Program of Tsinghua University; the fellowship of China Postdoctoral Science Foundation 2022M721875; and long lasting support of JBCA machine learning group, IAU OAD endorsed project team RGZ\_CN (2018) and EMU\_Zoo (2022). BDS acknowledges support from a UK Research and Innovation Future Leaders Fellowship [grant number MR/T044136/1].

This publication makes use of data products from the {\it Wide-field Infrared Survey Explorer}.  The {\it Wide-field Infrared Survey Explorer} is a joint project of the University of California, Los Angeles, and the Jet Propulsion Laboratory/California Institute of Technology, funded by the National Aeronautics and Space Administration.
This publication makes use of radio data from  the Karl G. Jansky Very Large Array (operated by NRAO).  The National Radio Astronomy Observatory is a facility of the National Science Foundation operated under cooperative agreement by Associated Universities, Inc.

\section*{Data Availability}
The RGZ DR1 (comprising of the 4 tables for the FIRST and ATLAS samples) is available for download as supplementary tables to this paper. Alternatively the DR1 tables can also be found at Zenodo (DOI: 10.5281/zenodo.14195049).

\bibliographystyle{mnras} 
\bibliography{mn-jour,paperef}

\appendix
\section{ATLAS subjects with low consensus}
\label{atlasappendix}
{\bf{This appendix presents a subset of the ATLAS subjects which have the lowest consensus level while containing only one source within the subject.}}

\begin{figure*}
\includegraphics[scale=.9]{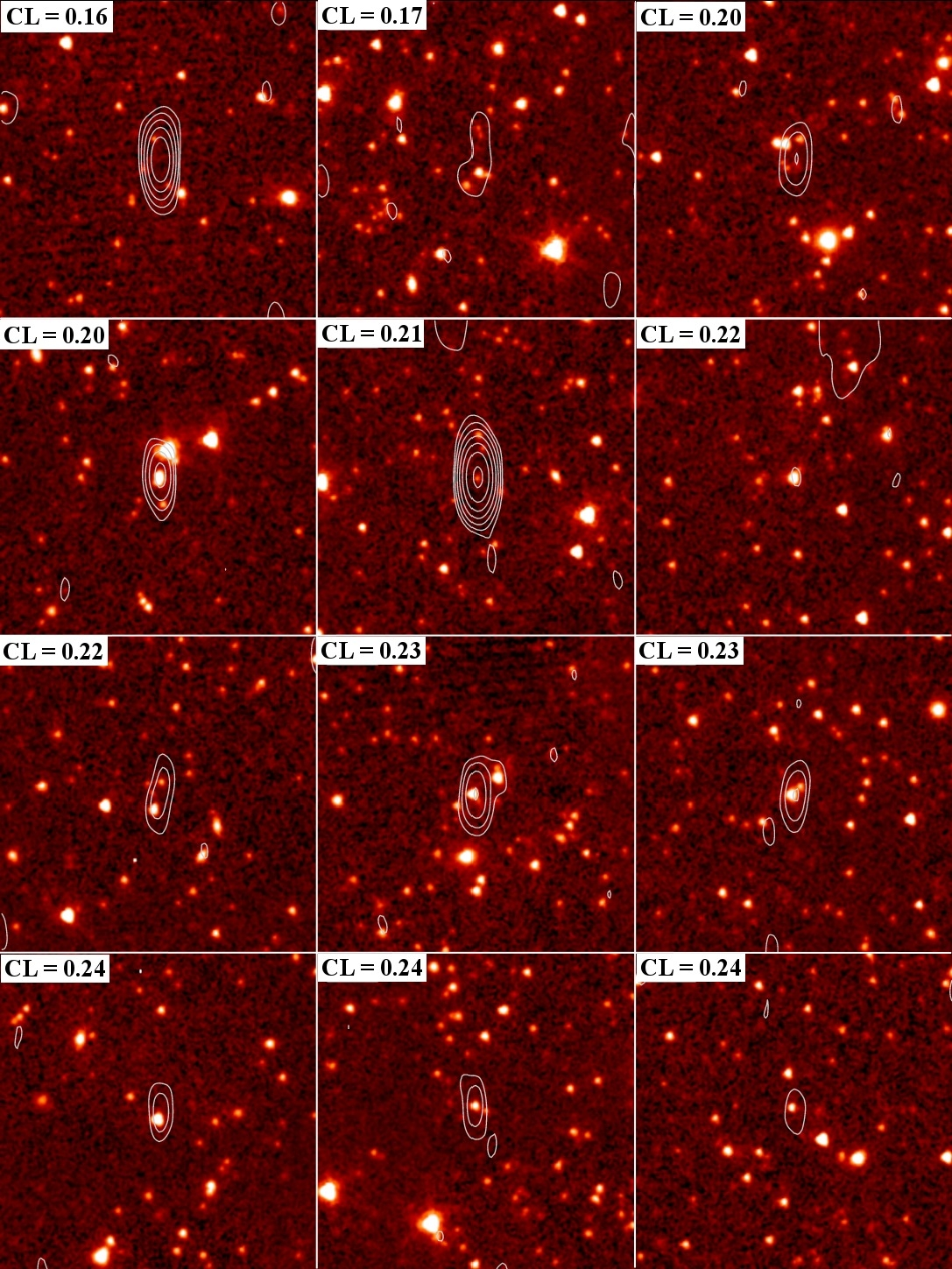}
\caption{Twelve RGZ subjects with the lowest consensus level for ATLAS subjects which only contain one radio source per subject. Each panel shows the ATLAS 1.4 GHz emission overlaid in white contours on the SWIRE 3.6~$\micron$m image in the background.  The weighted consensus level is marked in the top-left corner of each panel.}
\label{atlaslowcl}
\end{figure*}

\label{lastpage}
\end{document}